\documentstyle[12pt,a4]{article}
\input epsf

\global\arraycolsep=2pt

\begin{document}

\begin{titlepage}

\begin{flushright}
CERN-TH.7487/94\\
hep-ph/9412265
\end{flushright}

\vspace{0.5cm}

\begin{center}
\Large\bf Scale Setting in QCD and the\\
Momentum Flow in Feynman Diagrams
\end{center}

\vspace{1.0cm}

\begin{center}
Matthias Neubert\\
{\sl Theory Division, CERN, CH-1211 Geneva 23, Switzerland}
\end{center}

\vspace{1.2cm}

\begin{abstract}
We present a formalism to evaluate QCD diagrams with a single virtual
gluon using a running coupling constant at the vertices. This method,
which corresponds to an all-order resummation of certain terms in a
perturbative series, provides a description of the momentum flow
through the gluon propagator. It can be viewed as a generalization of
the scale-setting prescription of Brodsky, Lepage and Mackenzie to
all orders in perturbation theory. In particular, the approach can be
used to investigate why in some cases the ``typical'' momenta in a
loop diagram are different from the ``natural'' scale of the process.
It offers an intuitive understanding of the appearance of infrared
renormalons in perturbation theory and their connection to the rate
of convergence of a perturbative series. Moreover, it allows one to
separate short- and long-distance contributions by introducing a hard
factorization scale. Several applications to one- and two-scale
problems are discussed in detail.
\end{abstract}

\vspace{2.0cm}

\noindent
CERN-TH.7487/94\\
December 1994

\end{titlepage}

\section{Introduction}

Perturbative expansions in field theory are asymptotic, making it
necessary from a conceptual point of view (if not from a practical
one) to truncate any perturbative series at some finite order in the
coupling constant. This truncation introduces renormalization-scale
and scheme dependences. In QCD, this poses a problem because of the
strong scale dependence of the running coupling constant
$\alpha_s(\mu^2)$ in the low-momentum region. In many cases this puts
limitations on the precision with which experimental data can be
described in terms of the fundamental theory of strong interactions.

Several prescriptions on how to fix the renormalization scale and
scheme in a truncated perturbative series have been proposed
\cite{fix1}--\cite{fix8}. They all rely, in some way or another, on a
guess about the size of uncalculated higher-order contributions. This
guess can be based on criteria such as the apparent rate of
convergence of a series, the size of the coefficient of the last term
in the truncated series, or the sensitivity to changes of the
renormalization scale and scheme. It can also rely on physical
criteria such as the role of various mass scales in a given problem.
An example for such a physical scheme is provided by the
scale-setting prescription of Brodsky, Lepage and Mackenzie (BLM),
which amounts to absorb certain vacuum polarization effects appearing
at two-loop order into the one-loop running coupling constant
\cite{BLM}. Extensions of this scheme beyond the one-loop order have
been considered in Refs.~\cite{fix8}--\cite{LeMa}.

A large higher-order coefficient in a perturbative series can arise
from an anomalously large contribution of a particular set of
higher-order diagrams. For instance, one can imagine a series with a
large two-loop coefficient, but small one-loop and higher-order
coefficients. Such ``genuine'' higher-order effects are hard to
anticipate without a detailed calculation. On the other hand, it is
possible that large higher-order corrections result from an
inappropriate choice of the renormalization scale or scheme. In
particular, a change of the scale in the running coupling constant at
one-loop order leads to a change proportional to $\beta_0\,
\alpha_s^2$ at two-loop order, where $\beta_0 = 11-\frac{2}{3}\,n_f$
is the first coefficient of the $\beta$-function, and $n_f$ denotes
the number of light quark flavours. Since $\beta_0$ is large in QCD,
such effects can be numerically significant. An appropriate choice of
scale should try to minimize this type of higher-order terms.
Surprisingly, however, one finds that even in one-scale problems
using the ``natural'' scale in the running coupling constant at
one-loop order often leaves large corrections of order
$\beta_0\,\alpha_s^2$. Let us consider some examples related to heavy
quark systems: (i) the relation between the pole mass $m_b$ of the
bottom quark and the mass $\overline{m}_b(m_b)$ renormalized in the
modified minimal subtraction ($\overline{\rm MS}$) scheme; (ii) the
ratio of the decay constants of pseudoscalar and vector mesons, $B$
and $B^*$, in the so-called static limit; (iii) the parton model
prediction for the semi-leptonic decay rate $\Gamma(b\to u\,e\,
\bar\nu_e)$. At two-loop order, the corresponding perturbative series
are \cite{Gray}--\cite{LSW}
\begin{eqnarray}\label{ex1}
   {m_b\over\overline{m}_b(m_b)}
   &=& 1 + {4\over 3}\,{\alpha_s(m_b^2)\over\pi}
    + (1.56\beta_0 - 1.05)\,\Bigg( {\alpha_s(m_b^2)\over\pi}
    \Bigg)^2 + \dots \,, \nonumber\\
   {f_B^{\rm stat}\over f_{B^*}^{\rm stat}}
   &=& 1 + {2\over 3}\,{\alpha_s(m_b^2)\over\pi}
    + (0.53\beta_0 + 2.41)\,\Bigg( {\alpha_s(m_b^2)\over\pi}
    \Bigg)^2 + \dots \,, \nonumber\\
   {\Gamma(b\to u\,e\,\bar\nu_e)\over\Gamma_{\rm tree~level}}
   &=& 1 - 2.41\,{\alpha_s(m_b^2)\over\pi}
    - (3.22\beta_0 + k)\,\Bigg( {\alpha_s(m_b^2)\over\pi}
    \Bigg)^2 + \dots \,,
\end{eqnarray}
where the constant $k$ is yet unknown. In all cases the running
coupling constant is renormalized in the $\overline{\rm MS}$ scheme.
For $n_f=4$, the parts of the two-loop coefficients proportional to
$\beta_0$ are 13.02, 4.40, and 26.08, respectively. At least in the
first two examples these terms provide the dominant part of the
two-loop corrections. Moreover, in all three cases the two-loop
corrections are quite large, casting doubt on the convergence of the
perturbative series. An important example related to light quarks is
the so-called $D$-function, which is proportional to the derivative
of the correlator of two vector currents containing massless quark
fields. At two-loop order, one finds \cite{Ree1}--\cite{Ree3}
\begin{equation}\label{ex2}
   D(Q^2) = 1 + {\alpha_s(Q^2)\over\pi}
    + (0.17\beta_0 + 0.08)\,\Bigg( {\alpha_s(Q^2)\over\pi}
    \Bigg)^2 + \dots \,,
\end{equation}
where $Q^2$ denotes the euclidean momentum transfer. Again, the term
proportional to $\beta_0$ dominates the two-loop coefficient.
However, the absolute size of the two-loop correction is smaller than
in the cases considered above.

The appearance of large corrections proportional to
$\beta_0\,\alpha_s^2$ can be interpreted as an inappropriate choice
of scale in the one-loop running coupling constant. BLM have argued
that one should absorb these terms, which arise from self-energy
corrections to the gluon propagator, into the one-loop running
coupling constant. This prescription defines the so-called BLM scale
$\mu_{\rm BLM}$. For the above examples, one finds (in the
$\overline{\rm MS}$ scheme): $\mu_{\rm BLM}=0.10\,m_b$, $0.21\,m_b$,
$0.07\,m_b$, and $0.71\sqrt{Q^2}$. Clearly, if the BLM scale is to be
interpreted as a ``typical'' scale of virtual momenta in the
corresponding Feynman diagrams, the question arises why it is often
much lower than the ``natural'' mass scale in the problem at hand.

To analyse this issue, we propose a generalization of the BLM
prescription. Consider the perturbative calculation of a physical
(i.e.\ renormalization-scheme invariant and infrared finite) quantity
$S(M^2)$, which depends upon a single large mass scale $M$. We
restrict our discussion to Green functions without external gluons. A
generalization of the method to physical cross sections and inclusive
decay rates, which receive both virtual and real gluon corrections,
will be presented elsewhere \cite{part2}. The aim of the BLM
prescription is to ``guess'' the average virtuality of the gluon in a
loop diagram and to use it as the scale in the running coupling
constant. Clearly, a better way to proceed would be to perform the
calculation with a running coupling constant $\alpha_s(-k^2)$ at the
vertices, where $k$ is the momentum flowing through the virtual gluon
line. The result of such an improved calculation, which we denote by
$S_{\rm res}(M^2)$ since it corresponds to a partial resummation of
the perturbative series (see below), may be written as
\begin{equation}\label{resum}
   S_{\rm res}(M^2) = \int\!{\rm d}^4 k\,\alpha_s(-k^2)\,
   f(k,M,\dots)\equiv \int\limits_0^\infty\!{\rm d}t\,w(t)\,
   {\alpha_s(t M^2)\over 4\pi} \,,
\end{equation}
where $f(k,M,\dots)$ is the integrand of the Feynman diagrams. The
function $w(t)$ describes the distribution of virtualities in the
loop calculation. The integral over this distribution function
corresponds to an average of the running coupling constant over the
loop momenta in one-loop diagrams and, in a way, provides the optimal
improvement that can be achieved without a complete higher-order
calculation.

The fact that the integration in (\ref{resum}) extends to $t\to 0$
indicates the appearance of non-perturbative effects. It makes
explicit that any perturbative calculation receives long-distance
contributions from the integration over low momenta in Feynman
diagrams. We shall discuss the significance of these contributions at
length below. Here we just note that our approach goes beyond
perturbation theory since it is equivalent to the resummation of an
infinite number of terms in the perturbative series for the quantity
$S(M^2)$. Let us write this series in the general form
\begin{equation}\label{SM}
   S_{\rm pert}(M^2) = \sum_{n=1}^\infty \Bigg(
   {\alpha_s(M^2)\over 4\pi} \Bigg)^n\,S_n \,.
\end{equation}
The expansion coefficients $S_n$ can be written as a power series in
the number of light quark flavours or, equivalently, as a series in
powers of $\beta_0$. We restrict ourselves to cases where the
calculation of the one-loop coefficient $S_1$ does not involve the
non-abelian gluon self couplings, and where there are no external
gluons. Then the first dependence on $n_f$ comes at two-loop order,
and in general the coefficients can be written in the form
\begin{equation}\label{Snexp}
   S_n = c_n\,\beta_0^{n-1} + d_n\,\beta_0^{n-2} + \dots
   + k_n = c_n\,\Big( - \textstyle{2\over 3}\,n_f \Big)^{n-1}
   + O(n_f^{n-2}) \,.
\end{equation}
The coefficient $c_n$ in front of the highest power of $\beta_0$ is
proportional to the coefficient in front of the highest power of
$n_f$. This implies that $c_n$ can be calculated from ``quasi
one-loop'' diagrams, in which a gluon line is dressed by $(n-1)$
light-quark loops. It is the partial series built up by the terms
proportional to $c_n$ which is resummed in our approach. To see this,
we use the one-loop renormalization-group equation for the running
coupling constant to relate $\alpha_s(t M^2)$ in (\ref{resum}) to
$\alpha_s(M^2)$:
\begin{equation}\label{asrel}
   \alpha_s(t M^2) = \alpha_s(M^2)\,\sum_{n=1}^\infty
   \Bigg( {\beta_0\,\alpha_s(M^2)\over 4\pi} \Bigg)^{n-1}\,
   (-\ln t)^{n-1} \,.
\end{equation}
Under the above assumptions it is easy to see that the running of the
coupling constant is the only source of terms of order
$\beta_0^{n-1}\alpha_s^n$ in the series (\ref{SM}). We thus conclude
that
\begin{equation}
   S_{\rm res}(M^2) = \int\limits_0^\infty\!{\rm d}t\,w(t)\,
   {\alpha_s(t\,M^2)\over 4\pi} = \sum_{n=1}^\infty
   \Bigg( {\alpha_s(M^2)\over 4\pi} \Bigg)^n\,
   c_n\,\beta_0^{n-1} \,.
\end{equation}
The BLM prescription is to absorb the two-loop term $c_2\,\beta_0\,
\alpha_s^2$ into a redefinition of the scale used in the coupling
constant at one-loop order. It is equivalent to setting $w_{\rm
BLM}(t) = c_1\,\delta(t-e^{-c_2/c_1})$, thus choosing an average
virtuality $\mu_{\rm BLM}^2 = \exp(-c_2/c_1)\,M^2$ for the gluon.
Already at this point it is clear that this can only be a good
approximation if the distribution function $w(t)$ is narrow. If it is
wide, it is better to perform the integral in (\ref{resum}), which
resums all terms of the form $c_n\,\beta_0^{n-1}\alpha_s^n$.

Clearly, the resummation in (\ref{resum}) does not replace a full
higher-order calculation. For instance, ``genuine'' two-loop
corrections not related to the running of the coupling constant are
not taken into account. Nevertheless, our approach can be considered
as an optimal improvement of one-loop calculations, which takes into
account the full information contained in one-loop diagrams combined
with the running of the coupling constant. As such, we believe that
the construction of the distribution function $w(t)$ is an
interesting new concept, which provides information that goes beyond
what is contained in a low-order perturbative calculation.

In Sect.~\ref{sec:2} we generalize the above discussion in such a way
that the scale and scheme independence of the procedure become
apparent. In Sect.~\ref{sec:add} we discuss the asymptotic behaviour
of the distribution function and its relation to the ultraviolet and
infrared properties of the perturbative series. In particular, we
trace the appearance of infrared renormalons and discuss how to
separate short- and long-distance contributions by introducing a hard
factorization scale. In Sects.~\ref{sec:3} and \ref{sec:4} we
calculate the distribution function for several one- and two-scale
problems, among them most of the quantities considered in (\ref{ex1})
and (\ref{ex2}). We present a detailed numerical analysis, in which
we compare the results for the resummed series to low-order
calculations and investigate the relative size of short- and
long-distance contributions. Section~\ref{sec:6} contains a summary
and conclusions.

\section{Construction of the distribution function}
\label{sec:2}

We will now repeat the above argument in a slightly more general
form, which allows us to keep track of scale and scheme dependence.
Consider some dimensionless, infrared-safe and renormalization-scheme
invariant quantity $S(M^2,z)$, which depends upon some large mass
scale $M$ and, in the most general case, on a set of dimensionless
parameters $z$. For instance, in a two-scale problem we may choose
$M^2=m_1 m_2$ and $z=m_2/m_1$. Let us investigate the perturbative
series of $S(M^2,z)$ in powers of the coupling constant
$\alpha_s^{\rm R}(\mu^2)$ renormalized at some scale $\mu$ and in
some renormalization scheme R:
\begin{equation}\label{series}
   S_{\rm pert}(M^2,z) = \sum_{n=1}^\infty \Bigg(
   {\alpha_s^{\rm R}(\mu^2)\over 4\pi} \Bigg)^n\,
   S_n^{\rm R}(\mu,M,z) \,.
\end{equation}
The scale and scheme dependence in (\ref{series}) cancels between
the coefficient functions and the coupling constant. As in
(\ref{Snexp}), the coefficients can be expanded in powers of
$\beta_0$,
\begin{equation}
   S_n^{\rm R}(\mu,M,z)
   = c_n^{\rm R}(\mu,M,z)\,\beta_0^{n-1} + \dots \,,
\end{equation}
and we define a function $w_{\rm R}(t,\mu,M,z)$ such that
\begin{equation}\label{wRdef}
   S_{\rm res}(M^2,z) = \int\limits_0^\infty\!{\rm d}t\,
   w_{\rm R}(t,\mu,M,z)\,{\alpha_s^{\rm R}(t\,\mu^2)\over 4\pi}
   = \sum_{n=1}^\infty \Bigg( {\alpha_s^{\rm R}(\mu^2)\over 4\pi}
   \Bigg)^n\,c_n^{\rm R}(\mu,M,z)\,\beta_0^{n-1} \,.
\end{equation}
Using (\ref{asrel}) and the fact that, by construction, the
distribution function must be independent of $n_f$, one can derive a
relation between the moments of the distribution function and the
coefficients $c_n^{\rm R}$. It reads
\begin{equation}\label{moments}
   \int\limits_0^\infty\!{\rm d}t\,w_{\rm R}(t,\mu,M,z)\,
   (-\ln t)^{n-1} = c_n^{\rm R}(\mu,M,z) \,;\quad n\ge 1 \,.
\end{equation}
By inverting this relation one can in principle obtain the
distribution function from the knowledge of the set of coefficients
$\{c_n^{\rm R}\}$. This observation is crucial, as it implies that in
order to construct the distribution function it is sufficient to
consider the perturbative series (\ref{series}) in the fictitious
limit $\beta_0\to\infty$ (or $n_f\to -\infty$).

To proceed, it is convenient to construct a generating function for
the coefficients $c_n^{\rm R}$, which we define as
\begin{equation}\label{Borel}
   \widetilde S_{\rm R}(u,\mu,M,z) = \sum_{n=1}^\infty
   {u^{n-1}\over\Gamma(n)}\,c_n^{\rm R}(\mu,M,z) \,,
\end{equation}
so that
\begin{equation}
   c_n^{\rm R}(\mu,M,z) = \Bigg( {{\rm d}\over{\rm d}u}
   \Bigg)^{n-1} \widetilde S_{\rm R}(u,\mu,M,z)
   \Big|_{u=0} \,.
\end{equation}
This generating function is the Borel transform of the series
(\ref{series}) with respect to the inverse coupling constant
\cite{tHof}, in the limit $\beta_0\to\infty$.
Eq.~(\ref{Borel}) can be formally inverted to give
\begin{equation}\label{Laplace}
   S_{\rm res}(M^2,z) = {1\over\beta_0}\,\int\limits_0^\infty\!
   {\rm d}u\,\widetilde S_{\rm R}(u,\mu,M,z)\,\exp\Bigg(
   -{4\pi u\over\beta_0\,\alpha_s^{\rm R}(\mu^2)} \Bigg) \,.
\end{equation}
In cases where the integral exists, this equation defines the Borel
sum of the partial series $S_{\rm res}(M^2,z)$. In general, however,
the Borel transform has singularities on the real $u$-axis,
corresponding a factorial growth of the expansion coefficients
$c_n^{\rm R}$. Much of the non-perturbative structure of QCD can be
inferred from a study of the Borel transform
\cite{tHof}--\cite{Bene}. Its singularities on the negative axis
arise from the large-momentum region in Feynman diagrams and are
called ultraviolet renormalons. They are Borel summable and pose no
problem to performing the Laplace integral in (\ref{Laplace}).
Singularities on the positive axis arise from the low-momentum region
in Feynman diagrams and are called infrared renormalons. Their
presence leads to an ambiguity in the evaluation of the Laplace
integral. In the following section we will discuss how this ambiguity
is reflected in the integral over the distribution function in
(\ref{resum}).

The Borel transform defined in (\ref{Borel}) can be calculated by
evaluating one-loop diagrams in which the gluon propagator in Landau
gauge is replaced by \cite{Bene}
\begin{equation}
   D_{ab}^{\mu\nu}(k) = i\delta_{ab}\,\Bigg( {e^C\over\mu^2}
   \Bigg)^{-u}\,{k^\mu k^\nu - g^{\mu\nu} k^2\over(-k^2)^{2+u}} \,,
\end{equation}
where $C$ is a scheme-dependent constant related to the finite part
of a renormalized fermion-loop insertion on the gluon propagator.
Since we have assumed that the expansion coefficients $c_n^{\rm R}$
are dimensionless, it follows that the Borel transform can be
factorized as
\begin{equation}
   \widetilde S_{\rm R}(u,\mu,M,z) = \Bigg(
   {e^C M^2\over\mu^2} \Bigg)^{-u}\,\widehat S(u,z) \,,
\end{equation}
where the new function $\widehat S(u,z)$ is scale- and
scheme-independent. From (\ref{moments}) and (\ref{Borel}), we find
that the Borel transform can be expressed in terms of the
distribution function by the integral relation
\begin{equation}
   \widetilde S_{\rm R}(u,\mu,M,z) = \int\limits_0^\infty\!
   {\rm d}t\,w_{\rm R}(t,\mu,M,z)\,t^{-u} \,,
\end{equation}
which can be inverted to give
\begin{equation}
   w_{\rm R}(t,\mu,M,z) = {1\over 2\pi i t}
   \int\limits_{u_0-i\infty}^{u_0+i\infty}\!{\rm d}u\,
   \widehat S(u,z)\,\Bigg( {t\,\mu^2\over e^C M^2} \Bigg)^u \,.
\end{equation}
The choice of $u_0$ is arbitrary provided the integral exists. It
follows that the distribution density $w_R\,{\rm d}t$ depends on $t$,
$\mu$ and $M$ only in the combination
\begin{equation}\label{taudef}
   \tau = {t\,\mu^2\over e^C M^2} \,.
\end{equation}
If we introduce $\tau$ as a new variable and define a new scale- and
scheme-independent function $\widehat w(\tau,z)$ so that
\begin{eqnarray}\label{wSrel}
   \widehat w(\tau,z) &=& {1\over 2\pi i}
    \int\limits_{u_0-i\infty}^{u_0+i\infty}\!{\rm d}u\,
    \widehat S(u,z)\,\tau^{u-1} \,, \nonumber\\
   \widehat S(u,z) &=& \int\limits_0^\infty\!{\rm d}\tau\,
    \widehat w(\tau,z)\,\tau^{-u} \,, \nonumber\\
   c_n^{\rm R}(1,z) &=& \int\limits_0^\infty\!{\rm d}\tau\,
    \widehat w(\tau,z)\,(-C-\ln\tau)^{n-1} \,,
\end{eqnarray}
eq.~(\ref{wRdef}) can be written in a form that makes explicit
the renormalization-scheme invariance of the perturbative series:
\begin{equation}\label{whatint}
   S_{\rm res}(M^2,z) = \int\limits_0^\infty\!{\rm d}\tau\,
   \widehat w(\tau,z)\,{\alpha_s(\tau e^C M^2)\over 4\pi}
   = \sum_{n=1}^\infty \Bigg( {\alpha_s^{\rm R}(\mu^2)\over 4\pi}
   \Bigg)^n\,c_n^{\rm R}(\mu,M,z)\,\beta_0^{n-1} \,.
\end{equation}
Note that the scheme-dependence of the constant $C$ is such that the
value of the coupling constant $\alpha_s(e^C\mu^2)$ is
scheme-independent. This implies that the product $e^{-C/2}
\Lambda_{\rm QCD}$ is scheme-independent, where $\Lambda_{\rm QCD}$
is the scale parameter in the one-loop expression for the running
coupling constant. We note that $C=-5/3$ in the $\overline{\rm MS}$
scheme, $C=-5/3+\gamma-\ln 4\pi$ in the MS scheme, and $C=0$ in the
so-called V scheme \cite{BLM}.

As pointed out in the introduction, the integral over the
distribution function in (\ref{whatint}) is an improved one-loop
approximation to the quantity $S(M^2,z)$. It is instructive to
compare this approximation to the BLM scale-setting prescription. We
find
\begin{eqnarray}\label{Sappr}
   S_{\rm res}(M,z) &=& \int\limits_0^\infty\!{\rm d}\tau\,
    \widehat w(\tau,z)\,{\alpha_s(\tau e^C M^2)\over 4\pi}
    = N\,\Bigg\langle {\alpha_s(\tau e^C M^2)\over\pi}
    \Bigg\rangle \nonumber\\
   &=& N\,{\alpha_s(\mu_{\rm BLM}^2)\over\pi}\,\Bigg\{ 1
    + \Delta\,\Bigg( {\beta_0\,\alpha_s(\mu_{\rm BLM}^2)
    \over 4\pi} \Bigg)^2 + \dots \Bigg\} \,,
\end{eqnarray}
where
\begin{eqnarray}\label{Deltadef}
   N &=& {1\over 4}\,\int\limits_0^\infty\!{\rm d}\tau\,
    \widehat w(\tau,z) \,, \nonumber\\
   \mu_{\rm BLM}^2 &=& \exp\Big( \langle \ln\tau \rangle + C
    \Big)\,M^2 \,, \nonumber\\
   \phantom{ \Bigg[ }
   \Delta &=& \sigma_\tau^2 = \langle \ln^2\!\tau \rangle
    - \langle \ln\tau \rangle^2 \,.
\end{eqnarray}
We use the symbol
\begin{equation}
   \langle f(\tau) \rangle =
   {\int\limits_0^\infty\!{\rm d}\tau\,\widehat w(\tau,z)\,
    f(\tau) \over \int\limits_0^\infty\!{\rm d}\tau\,
    \widehat w(\tau,z)}
\end{equation}
for the average of a function $f(\tau)$ over the distribution
$\widehat w(\tau,z)$. Both the value of the coupling constant
$\alpha_s(\mu_{\rm BLM}^2)$ and the parameter $\Delta$ are
renormalization-scheme invariant. We observe that the first
correction to the BLM scheme is related to the width of the
distribution function. Note, however, that in some cases the
distribution function is not of a definite sign and has no
probabilistic interpretation. Thus, it may happen that
$\sigma_\tau^2$ is negative or that the normalization integral $N$
vanishes, in which case $\langle f(\tau)\rangle$ would be
ill-defined. If the distribution function has a definite sign, on the
other hand, the quantity
\begin{equation}\label{deldef}
   \delta_{\rm BLM} = \Delta\,\Bigg( {\beta_0\,
   \alpha_s(\mu_{\rm BLM}^2)\over 4\pi} \Bigg)^2
   = {\langle \ln^2\!\tau \rangle - \langle \ln\tau \rangle^2
   \over \big[ \ln(M^2/\Lambda_{\rm V}^2) + \langle \ln\tau \rangle
   \big]^2} \,,
\end{equation}
provides a measure of the rate of convergence of the perturbative
series. Here
\begin{equation}
    \Lambda_{\rm V} = e^{-C/2}\,\Lambda_{\rm QCD}
    = e^{5/6}\,\Lambda_{\overline{\rm MS}}
\end{equation}
is a scheme-independent parameter, which coincides with the QCD scale
parameter in the V scheme.

At this point it is worthwhile to point out the advantages of our
resummation over the BLM prescription. In our approach all terms of
order $\beta_0^{n-1}\alpha_s^n$ in the perturbative series are
resummed exactly, whereas the BLM scheme only resums the two-loop
term of order $\beta_0\,\alpha_s^2$ correctly. Moreover, our scheme
is additive in the sense that the distribution function for the sum
of two quantities is the sum of the individual distribution
functions: $\widehat w_{A+B}= \widehat w_A+\widehat w_B$. No such
relation exists for the corresponding BLM scales. In particular, both
$A$ and $B$ can have small BLM scales, but the BLM scale for the sum
$(A+B)$ can be large, or vice versa. This brings us to the most
important point, which is that the size of the BLM scale cannot
always be used as an indicator for the rate of convergence of a
perturbative series. The size of higher-order coefficients depends on
the size of moments of the distribution function, not only on its
central value, which determines the BLM scale.

\section{Asymptotic behaviour, infrared renormalons and the
separation of long-distance contributions}
\label{sec:add}

Before we calculate the distribution function for specific examples
we address the relevance of its asymptotic behaviour for large and
small values of $\tau$. Clearly, the behaviour for $\tau\to\infty$ is
related to the ultraviolet properties of the perturbative series. In
particular, only if $\widehat w(\tau,z)$ vanishes faster than
$1/\tau$
the integral in (\ref{Sappr}) converges. Otherwise, our approach
provides the framework for a consistent cutoff regularization of
the series. Performing the integral up to a value $\tau_{\rm UV}$
corresponds to a hard momentum cutoff $\Lambda_{\rm UV}^2=\tau_{\rm
UV} M^2$.

More subtle is the infrared region $\tau\to 0$. As long as one stays
within perturbation theory one faces the problem that the integral
over the distribution function runs over the Landau pole in the
running coupling constant. One is forced to specify how to treat this
pole, for instance by deforming the integration contour. In general,
we may write
\begin{equation}
   {\beta_0\,\alpha_s(\tau e^C M^2)\over 4\pi}
   = {1\over\ln M^2/\Lambda_V^2 + \ln\tau} \nonumber\\
   = {\rm P}\,\bigg( {1\over\ln\tau-\ln\tau_L} \bigg)
   + \eta\,\delta(\ln\tau-\ln\tau_L) \,,
\end{equation}
where $\tau_L=\Lambda_V^2/M^2$ is the position of the Landau pole,
``P'' denotes the principle value, and $\eta$ is a complex parameter
of order unity, which depends on the regularization prescription.
This prescription dependence leads to an intrinsic ambiguity in the
perturbative definition of $S(M^2,z)$, reflecting the fact that in
(\ref{whatint}) we are trying to sum up a series which is not Borel
summable. This is how infrared renormalons make their appearance in
our approach. Following Refs.~\cite{BBren}--\cite{Chris}, we define
the renormalon ambiguity $\Delta S_{\rm ren}$ in the value of
$S(M^2,z)$ as the coefficient of $\eta$ and find
\begin{equation}\label{Drenorm}
   \Delta S_{\rm ren} = {\tau_L\over\beta_0}\,
   \widehat w(\tau_L,z) \simeq {w_0(z)\over\beta_0}\,
   \bigg( {\Lambda_V\over M} \bigg)^{2 k} \,.
\end{equation}
In the last step we have used the fact that $\tau_L\ll 1$ to expand
the distribution function:
\begin{equation}\label{wtauexp}
   \widehat w(\tau,z) = w_0(z)\,\tau^{k-1} + \dots \quad
   \mbox{for $\tau\to 0$.}
\end{equation}
It is thus the asymptotic behaviour of the distribution function that
determines the size of the renormalon ambiguity. Note that $k>0$ in
order for the integral over the distribution function to be infrared
convergent. We will see in examples that the power $k$ is related to
the position of the nearest infrared renormalon pole in the Borel
transform $\widehat S(u,z)$, which is located at $u_{\rm IR}=k$ in
the Borel plane.

The appearance of infrared renormalons acts as a reminder that in
(\ref{Sappr}) one is using perturbation theory in a regime where it
is known to break down, namely in the infrared region. Hence, the
result of any perturbative calculation in QCD is incomplete; it must
be supplemented by non-perturbative corrections. Only the sum of all
perturbative and non-perturbative contributions is unambiguous.
Unlike any finite-order calculation, the representation (\ref{Sappr})
makes explicit that perturbative calculations contain long-distance
contributions from the region of low momenta in Feynman diagrams.
Moreover, it provides a convenient way to separate these
long-distance contributions from the short-distance ones by
introducing a hard factorization scale $\lambda$. Thus, our approach
can be used to implement Wilson's construction of the operator
product expansion (OPE) \cite{Wils} in a literal way. Let us recall
that the OPE is not designed to separate perturbative and
non-perturbative effects, but to disentangle the physics on different
length scales. In the calculation of short-distance corrections (the
Wilson coefficient functions) one eliminates the contributions from
small virtualities ($k<\lambda$) in Feynman diagrams. These
long-distance contributions are attributed to some matrix elements of
higher-dimensional operators. Thus, we should write
\begin{eqnarray}\label{separ}
   S_{\rm res}(M^2,z) &=&
    \int\limits_{\lambda^2/M^2}^\infty\!{\rm d}\tau\,
    \widehat w(\tau,z)\,{\alpha_s(\tau e^C M^2)\over 4\pi}
    + \int\limits_0^{\lambda^2/M^2}\!{\rm d}\tau\,
    \widehat w(\tau,z)\,{\alpha_s(\tau e^C M^2)\over 4\pi}
    \nonumber\\
   \phantom{ \bigg[ }
   &\equiv& S_{\rm sd}(M^2,\lambda,z) + S_{\rm ld}(M^2,\lambda,z) \,.
\end{eqnarray}
Note that the factorization point $\tau=\lambda^2/M^2$ corresponds to
a scheme-dependent scale $\mu^2=e^C\lambda^2$ in the running coupling
constant, with $\mu=\lambda$ in the V scheme. The value of
$\alpha_s(e^C\lambda^2)$ is scheme-independent, however. As long as
$\lambda$ is chosen large enough, the short-distance contribution
$S_{\rm sd}$ can be reliably calculated in perturbation theory and is
free of renormalon ambiguities. The long-distance contribution
$S_{\rm ld}$ must be combined with other non-perturbative
corrections. Only the sum of all long-distance contributions is well
defined. Of course, the dependence on the arbitrary scale $\lambda$
must cancel in the final result. This $\lambda$-dependence can be
controlled in perturbation theory by means of the
renormalization-group equation
\begin{equation}
   \lambda\,{{\rm d}\over{\rm d}\lambda}\,S_{\rm sd}(M^2,\lambda,z)
   = -\lambda\,{{\rm d}\over{\rm d}\lambda}\,
   S_{\rm ld}(M^2,\lambda,z)
   = -{\alpha_s(e^C \lambda^2)\over 2\pi}\,{\lambda^2\over M^2}\,
   \widehat w(\lambda^2/M^2,z) \,.
\end{equation}

Since we have required that the quantity $S(M^2,z)$ be infrared
safe, the long-distance contribution $S_{\rm ld}$ is finite, and it
is usually assumed that it is small compared to the short-distance
contribution. Eq.~(\ref{separ}) allows us to quantify this statement.
If the factorization scale is chosen such that $\lambda\ll M$, the
$M$-dependence of the long-distance contribution is again determined
by the asymptotic behaviour of the distribution function for small
values of $\tau$. We find
\begin{equation}\label{DIR}
   S_{\rm ld}(M^2,\lambda,z) \simeq {w_0(z)\over M^{2k}}\,
   \int\limits_0^{\lambda^2}\!{\rm d}\mu^2\,\mu^{2(k-1)}\,
   {\alpha_s(e^C\mu^2)\over 4\pi} \equiv \pm\Bigg(
   {\Lambda(\lambda)\over M} \Bigg)^{2 k} \,,
\end{equation}
where $\Lambda(\lambda)$ is of order the QCD scale, and the sign is
determined by the sign of $w_0$. The long-distance contribution is
indeed parametrically small, suppressed by inverse powers of the
large mass scale $M$. It is, of course, no accident that
long-distance effects appear at the same order in $1/M$ as infrared
renormalon ambiguities. Both effects are exponentially small in the
coupling constant and thus not seen in any finite-order perturbative
calculation. Nevertheless, we will see that in cases where there is a
nearby infrared renormalon (i.e.\ when the power $k$ is small) the
long-distance contribution and the renormalon ambiguity can be
numerically significant. One can try to estimate the size of $S_{\rm
ld}$ by incorporating certain non-perturbative effects into the
integral over the distribution function, for instance by using a more
realistic ansatz for the running coupling constant in the infrared
region. For instance, one may assume that the coupling constant
$\alpha_s(\mu^2)$ stays positive and approaches a constant for
$\mu\to 0$. It is then possible to perform the $\tau$-integral
without encountering a Landau pole. This yields to an unambiguous
(though model-dependent) result for the long-distance contribution.

\section{Heavy quark systems}
\label{sec:3}

We now illustrate the formalism developed above with some quantities
related to heavy quarks, which provide prototype examples for
large-scale problems in QCD. The large mass scale $M$ is provided by
a heavy quark mass $m_Q$. Sects.~\ref{sec:3.1}--\ref{sec:3.3} deal
with the derivation of the distribution function for several
quantities of interest. In Sect.~\ref{sec:3.4} we present a numerical
analysis of the results.

\subsection{Pole mass of a heavy quark}
\label{sec:3.1}

We start with the relation between the pole mass $m_Q$ and the
(infinite) bare mass $m_0$, which appears as a parameter in the QCD
Lagrangian. We define
\begin{equation}
   S_m(m_Q^2) = {m_Q\over m_0} - 1
\end{equation}
and aim for a representation of this quantity as an integral over a
distribution function $\widehat w_m(\tau)$, as shown in
(\ref{Sappr}). In the large-$\beta_0$ limit, the Borel transform
corresponding to the perturbative series for $S_m(m_Q^2)$ has the
simple form \cite{BBren}
\begin{equation}\label{Smu}
   \widehat S_m(u) = 6 C_F\,(1-u)\,
   {\Gamma(u)\,\Gamma(1-2 u)\over\Gamma(3-u)} \,,
\end{equation}
where $C_F=4/3$ is a colour factor. To derive the distribution
function, we rewrite
\begin{equation}
   (1-u)\,{\Gamma(u)\,\Gamma(1-2 u)\over\Gamma(3-u)}
    = \int\limits_0^1\!{\rm d}x \int\limits_0^1\!{\rm d}y\,
    x^{u-1} (1-x)^{-2 u}\,y^{1-u}
\end{equation}
and use the second relation in (\ref{wSrel}) to obtain
\begin{equation}
   \widehat w_m(\tau) = 6 C_F \int\limits_0^1\!{{\rm d}x\over x}
   \int\limits_0^1\!{\rm d}y\,y\,
   \delta\bigg( \tau - {(1-x)^2 y\over x} \bigg) \,.
\end{equation}
After a straightforward calculation, we find
\begin{equation}
   \widehat w_m(\tau) = C_F\,\Bigg\{ {\tau\over 2}
   + \bigg( 1 - {\tau\over 2} \bigg)\,\sqrt{1 + {4\over\tau}}
   \Bigg\} \,.
\end{equation}
The small-$\tau$ behaviour of $\widehat w_m(\tau)$ is
\begin{equation}\label{wmzero}
   \widehat w_m(\tau) = {2 C_F\over\sqrt{\tau}} + O(\sqrt{\tau}) \,,
\end{equation}
corresponding to $k=1/2$ in (\ref{wtauexp}). This behaviour is
associated with the infrared renormalon pole at $u=1/2$ in the Borel
transform in (\ref{Smu}). According to (\ref{Drenorm}), the
corresponding ambiguity in the perturbative series for $m_Q/m_0$ is
\begin{equation}\label{dmren}
   (\Delta S_m)_{\rm ren} = {(\Delta m_Q)_{\rm ren}\over m_Q}
   = {8\over 3\beta_0}\,{\Lambda_V\over m_Q} \,,
\end{equation}
implying an ambiguity $(\Delta m_Q)_{\rm ren}=(8/3\beta_0)\,
\Lambda_V$ in the value of the pole mass \cite{BBren,Bigiren}.

For large values of $\tau$ the distribution function decreases as
$3 C_F/\tau$, so that the integral in (\ref{Sappr}) is
logarithmically divergent. The divergence is removed by a
renormalization of the bare mass. It turns out that the calculation
of the distribution function is complicated if one chooses the
$\overline{\rm MS}$ scheme for this purpose. The Borel transform
corresponding to the ratio $m_Q/\overline{m}_Q(m_Q)$ is given by
\cite{BBren}
\begin{equation}\label{SmBorel}
   \widehat S_m^{\overline{\rm MS}}(u) = C_F\,\Bigg\{
   6 (1-u)\,{\Gamma(u)\,\Gamma(1-2 u)\over\Gamma(3-u)}
   + e^{-5 u/3}\,\bigg( -{3\over u} + R(u) \bigg) \Bigg\} \,,
\end{equation}
where
\begin{equation}
   R(u) = - {5\over 2} + {35\over 24}\,u
   + \bigg( \zeta(3) - {83\over 144} \bigg)\,u^2 + \dots
\end{equation}
is a rather complicated function. Instead, we find it instructive to
consider a class of renormalization schemes R$[r]$ which are more
convenient for our calculation. We define the Borel transform
corresponding to the ratio $m_Q/m_Q^{\rm R}$ as
\begin{equation}
   \widehat S_m^{\rm R}(u) = C_F\,\Bigg\{
   6 (1-u)\,{\Gamma(u)\,\Gamma(1-2 u)\over\Gamma(3-u)}
   - {3\over u}\,e^{-r u} \Bigg\}
\end{equation}
and treat $r$ as a free parameter. From a comparison of the Borel
transforms $\widehat S_m^{\overline{\rm MS}}$ and $\widehat
S_m^{\rm R}$ one finds that the relation between the so-defined mass
$m_Q^{\rm R}$ and the mass renormalized in the $\overline{\rm MS}$
scheme is given by
\begin{eqnarray}\label{massrel}
   {m_Q^{\rm R}\over\overline{m}_Q(m_Q)} &=& 1
    - \bigg( r - {5\over 6} \bigg)\,{\bar\alpha_s\over\pi}
    + {\beta_0\over 8}\,\bigg( r^2 - {10 r\over 3} + {15\over 4}
    \bigg)\,\bigg( {\bar\alpha_s\over\pi} \bigg)^2 \nonumber\\
   &&\mbox{}- {\beta_0^2\over 48}\,\bigg( r^3 - 5 r^2
    + {25 r\over 3} - {751\over 216} - 2\zeta(3) \bigg)\,
    \bigg( {\bar\alpha_s\over\pi} \bigg)^3 + \dots \,,
\end{eqnarray}
where $\bar\alpha_s=\alpha_s(m_Q^2)$ denotes the coupling constant in
the $\overline{\rm MS}$ scheme. It is a simple exercise to calculate
the distribution function for the ratio $m_Q/m_Q^{\rm R}$. The result
is
\begin{equation}
   \widehat w_m^{\rm R}(\tau) = \widehat w_m(\tau)
   - {3 C_F\over\tau}\,\Theta(\tau-e^r) \,,
\end{equation}
which simply amounts to a subtraction of the high-momentum
contributions, leaving the low-momentum region unaffected. The
subtracted distribution function falls off as $1/\tau^2$ for large
values of $\tau$, so that the integral over $\tau$ is convergent.

With the distribution function $\widehat w_m^{\rm R}(\tau)$ we
compute
\begin{eqnarray}
   N &=& \bigg( {3\over 8} + {3 r\over 4} \bigg)\,C_F
    = {1\over 2} + r \,, \nonumber\\
   \langle \ln\tau \rangle &=& {r^2 - {1\over 2} - {2\over 3}\,\pi^2
    \over 2r + 1} \,, \nonumber\\
   \langle \ln^2\!\tau \rangle &=& {2\over 3}\,
    {r^3 + {3\over 4} + 12\zeta(3) + \pi^2\over 2r + 1} \,,
\end{eqnarray}
which is all one needs to calculate the BLM scale and the parameter
$\Delta$ defined in (\ref{Deltadef}). Two cases are particularly
interesting:
\begin{eqnarray}
   \mbox{scheme R1:} \qquad
   r &=& {5\over 6} \,, \nonumber\\
   \mbox{scheme R2:} \qquad
   r &=& r_0 = \sqrt{{1\over 2} + {2\pi^2\over 3}} \simeq 2.661 \,.
\end{eqnarray}
Since the one-loop coefficient in (\ref{massrel}) vanishes for
$r=5/6$, scheme R1 is similar to the $\overline{\rm MS}$ scheme. The
relation between the mass definitions in the two schemes is
\begin{equation}
   {m_Q^{\rm R1}\over\overline{m}_Q(m_Q)} = 1
   + 0.208\,\beta_0\bigg( {\bar\alpha_s\over\pi} \bigg)^2
   + 0.038\,\beta_0^2\,\bigg( {\bar\alpha_s\over\pi} \bigg)^3
   + \dots \,,
\end{equation}
which is a nicely converging series. For the bottom quark the ratio
equals 1.009. The scheme R2 is chosen such that $\langle\ln\tau
\rangle=0$, so that the BLM scale is given by $\mu_{\rm BLM}^2 = e^C
m_Q^2$. In this case one finds
\begin{equation}
   {m_Q^{\rm R2}\over\overline{m}_Q(m_Q)} = 1
   - 1.827\,{\bar\alpha_s\over\pi}
   + 0.245\,\beta_0\bigg( {\bar\alpha_s\over\pi} \bigg)^2
   + 0.006\,\beta_0^2\,\bigg( {\bar\alpha_s\over\pi} \bigg)^3
   + \dots \,,
\end{equation}
which has a sizeable one-loop coefficient but is again nicely
converging. For the bottom quark the ratio equals 0.883.

\begin{table}[t]
\centerline{\parbox{15cm}{\caption{\label{tab:1}
Parameters related to the distribution function $\widehat w_m^{\rm
R}(\tau)$ for three different mass renormalization schemes. We use
the one-loop expression for the running coupling constant in the
$\overline{\rm MS}$ scheme, normalized such that
$\alpha_s(m_b^2)=0.218$ for $m_b=4.8$ GeV.}}}
\vspace{0.5cm}
\centerline{\begin{tabular}{l|ccccc}
\hline\hline
\rule[-0.2cm]{0cm}{0.7cm} & $\mu_{\rm BLM}^V$ &
 $\mu_{\rm BLM}^{\overline{\rm MS}}$ & $\alpha_s(\mu_{\rm BLM}^2)$
 & $\Delta$ & $\delta_{\rm BLM}$ \\
\hline
\rule[-0.1cm]{0cm}{0.6cm} $m_b/m_b^{\rm R1}$ & $0.302\,m_b$ &
 $0.131\,m_b$ & 0.55 & 0.672 & 0.10 \\
\rule[-0.1cm]{0cm}{0.6cm} $m_b/m_b^{\rm R2}$ & $m_b$ &
 $0.435\,m_b$ & 0.29 & 4.628 & 0.17 \\
\rule[-0.1cm]{0cm}{0.6cm} $m_b/\overline{m}_b(m_b)$ &
 $0.221\,m_b$ & $0.096\,m_b$ & 0.73 & $-4.337\phantom{-}$ &
 $-1.19\phantom{-}$ \\
\hline\hline
\end{tabular}}
\vspace{0.5cm}
\end{table}

\begin{figure}[htb]
   \vspace{0.5cm}
   \epsfysize=6cm
   \centerline{\epsffile{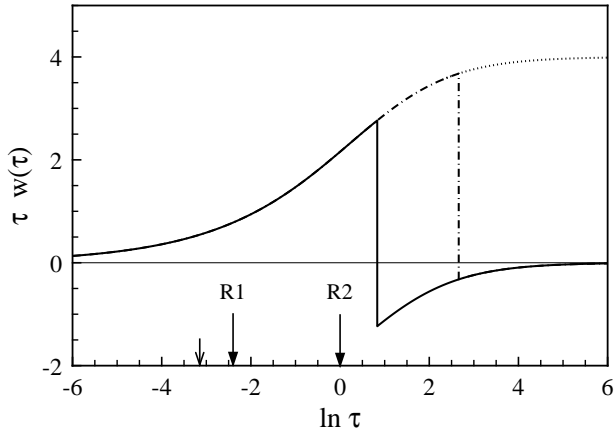}}
   \centerline{\parbox{13cm}{\caption{\label{fig:1}
Distribution function for the ratio $m_b/m_b^{\rm R}$ in the two
renormalization schemes R1 (solid line) and R2 (dashed-dotted line).
The dotted line shows the unsubtracted distribution function for the
ratio of the pole mass and the bare mass. The long arrows indicate
the position of the BLM scale in the schemes R1 and R2. The small
arrow shows the factorization point, which separates short- and
long-distance contributions.}}}
\end{figure}

In Table~\ref{tab:1} we give, for the case of the bottom quark, the
results for the BLM scale in three different mass renormalization
schemes (R1, R2, $\overline{\rm MS}$) and for two renormalization
schemes for the coupling constant (V and $\overline{\rm MS}$). We
also quote the corresponding scheme-independent values of
$\alpha_s(\mu_{\rm BLM}^2)$, as well as the parameters $\Delta$ and
$\delta_{\rm BLM}$ defined in (\ref{Deltadef}) and (\ref{deldef}). We
note that the distribution function corresponding to the
$\overline{\rm MS}$ scheme must satisfy
\begin{eqnarray}
   N &=& C_F = {4\over 3} \,, \nonumber\\
   \langle \ln\tau \rangle &=& - {53\over 96} - {\pi^2\over 4}
    \simeq -3.019 \,, \nonumber\\
   \langle \ln^2\!\tau \rangle &=& {7\over 2}\,\zeta(3)
    - {1637\over 864} + {\pi^2\over 4} \simeq 4.780 \,,
\end{eqnarray}
as can be derived from an expansion of the Borel transform
(\ref{SmBorel}) in powers of $u$. The large negative values of the
parameters $\Delta$ and $\delta_{\rm BLM}$ for the ratio
$m_b/\overline{m}_b(m_b)$ indicate that there are large corrections
to the BLM scheme which decrease the value of the perturbative
series. In other words, the BLM scale is too low and does not really
represent an ``average'' virtuality. The reason for the strong
dependence of the BLM scale on the subtraction scheme becomes
apparent from the shape of the distribution functions shown in
Fig.~\ref{fig:1}. We find it most useful to show the product
$\tau\,\widehat w(\tau)$ as a function of $\ln\tau$, since then the
integrals $\langle\ln^n\!\tau\rangle$ have a direct graphical
interpretation. The long arrows indicate the position of the BLM
scale in the schemes R1 and R2. The small arrow shows the point
$\tau=\lambda^2/m_b^2$ for $\lambda=1$ GeV, which will later be used
to separate short- and long-distance contributions
[cf.~(\ref{separ})]. In order to associate mass scales with the
$\tau$-values in the figure, we note that
\begin{equation}
   \ln{\mu^2\over m_b^2} = \ln\tau + C \,,
\end{equation}
where $\mu$ is the scale in the running coupling constant. In the V
scheme, where $C=0$, the point $\tau=\lambda^2/m_b^2$ corresponds to
the scale $\mu=\lambda$, whereas $\ln\tau=0$ corresponds to
$\mu=m_b$. In the $\overline{\rm MS}$ scheme, where $C=-5/3$, the
point $\mu=m_b$ corresponds to $\ln\tau=5/3$.

We observe that the pole mass gets contributions from all momentum
scales, and it is only the subtraction of the high-momentum tail that
leads to a negative value of $\langle\ln\tau\rangle$ in the scheme R1
(and similar for the $\overline{\rm MS}$ scheme). If the subtraction
point is chosen as low as in R1, significant cancellations take place
between positive and negative contributions in the integral over the
distribution function. The results is a small one-loop coefficient
$N$ in (\ref{Deltadef}), yielding in turn a small BLM scale. Thus,
the interpretation of the BLM scale as a ``typical'' scale in a
process becomes misleading if the distribution function gets
contributions of opposite sign. Note that the situation encountered
here is generic for quantities which require a subtraction of
ultraviolet divergences. In such cases, a low value of the BLM scale
does not necessarily imply a bad convergence of the perturbative
series.

\begin{figure}[htb]
   \vspace{0.5cm}
   \epsfysize=6cm
   \centerline{\epsffile{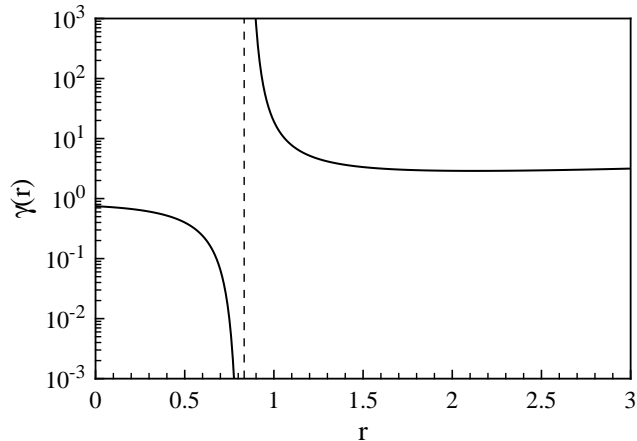}}
   \centerline{\parbox{13cm}{\caption{\label{fig:extra}
The function $\gamma(r)$ defined in (\protect\ref{gamdef}).}}}
\end{figure}

The series in (\ref{massrel}) is interesting by itself. Since both
$m_Q^{\rm R}$ and $\overline{m}_Q(m_Q)$ are subtracted at a large
mass scale, their ratio has a well-behaved expansion in powers of
$\alpha_s(m_Q^2)$ as long as $r$ is of order unity. In fact, for the
two choices of $r$ discussed above the series was rapidly converging.
Nevertheless, the BLM scale corresponding to this series exhibits a
strong dependence on $r$. We find $\mu_{\rm BLM}^2 = e^C\,
[\gamma(r)\,m_Q]^2$ with
\begin{equation}\label{gamdef}
   \ln\gamma(r) = {r^2 + {35\over 36}\over 4 r - {10\over 3}} \,.
\end{equation}
This function is shown in Fig.~\ref{fig:extra}. The reason for the
dramatic scheme dependence of $\mu_{\rm BLM}$ is simply that the
one-loop coefficient vanishes for $r=5/6$. This example shows that a
small BLM scale is not always related to a badly converging series.
It also shows the disadvantage of non-additivity of the BLM
prescription: As shown in Table~\ref{tab:1}, the BLM scale for the
ratio $m_Q/\overline{m}_Q(m_Q)$ is rather low. However, the same
ratio can be obtained by combining the two series for $m_Q/m_Q^{\rm
R}$ and $m_Q^{\rm R}/\overline{m}_Q(m_Q)$ in the scheme R2, both of
which have a much larger BLM scale. On the other hand, calculating
$m_Q/\overline{m}_Q(m_Q)$ by combining integrals over the appropriate
distribution functions one obtains a unique result, in which the
contributions from all scales are properly taken into account.

\subsection{Matching coefficients of heavy--light currents}
\label{sec:3.2}

A convenient way to analyse hadronic matrix elements of the
flavour-changing weak current $j^\mu = \bar q\,\gamma^\mu
(1-\gamma_5)\,Q$ between a hadron containing the heavy quark $Q$ and
some light final state is to go over to an effective theory, the
so-called heavy quark effective theory \cite{Geor}, in which such
matrix elements are systematically expanded in powers of $1/m_Q$.
When QCD is matched onto the effective theory, the current gets
replaced by \cite{review}
\begin{equation}
   j^\mu \to C_1(m_Q^2)\,\bar q\,\gamma^\mu (1-\gamma_5)\,h_v
   + C_2(m_Q^2)\,\bar q\,v^\mu (1+\gamma_5)\,h_v + O(1/m_Q) \,,
\end{equation}
where $v$ is the four-velocity of the hadron that contains the heavy
quark, and $h_v$ is the velocity-dependent heavy quark field in the
effective theory. The above form of the currents is correct if one
uses a regularization scheme with anti-commuting $\gamma_5$. The
matching coefficients $C_i(m_Q^2)$ can be calculated in perturbation
theory by comparing quark matrix elements of the currents in QCD and
in the effective theory. We define
\begin{equation}
   S_1(m_Q^2) = C_1(m_Q^2) - 1 \,,\qquad
   S_2(m_Q^2) = C_2(m_Q^2) \,.
\end{equation}
In the large-$\beta_0$ limit, the Borel transforms of the
perturbative series for these quantities are given by \cite{Chris}
\begin{eqnarray}\label{S1S2}
   \widehat S_1(u) &=& C_F\,( 3 u^2 - u - 3)\,
    {\Gamma(u)\,\Gamma(1-2u)\over\Gamma(3-u)} \,, \nonumber\\
   \widehat S_2(u) &=& 4 C_F\,
    {\Gamma(1+u)\,\Gamma(1-2u)\over\Gamma(3-u)} \,.
\end{eqnarray}
The corresponding distribution functions can be calculated as
outlined in the previous section. We find
\begin{eqnarray}\label{w1w2}
   \widehat w_1(\tau) &=& {C_F\over 2}\,\Bigg\{ 1
    - {7\over 6}\,\tau - {3\over 2\sqrt{1 + 4/\tau}}
    - {11-7\tau\over 6}\,\sqrt{1 + {4\over\tau}} \Bigg\} \,,
    \nonumber\\
   \widehat w_2(\tau) &=& {2 C_F\over 3}\,\Bigg\{
    (1+\tau)\,\sqrt{1 + {4\over\tau}} - 3 - \tau \Bigg\} \,.
\end{eqnarray}
The asymptotic behaviour for small values of $\tau$ is
\begin{equation}
   \widehat w_1(\tau) = - {11 C_F\over 6\sqrt{\tau}} + O(1)
    \,, \qquad
   \widehat w_2(\tau) = {4 C_F\over 3\sqrt{\tau}} + O(1) \,,
\end{equation}
corresponding to the infrared renormalon poles at $u=1/2$ in the
Borel transforms in (\ref{S1S2}). If we relate the corresponding
renormalon ambiguities to the ambiguity in the value of the pole mass
[cf.~(\ref{wmzero}) and (\ref{dmren})], we recover the relations
\begin{equation}
   (\Delta C_1)_{\rm ren} = - {11\over 12}\,
   {(\Delta m_Q)_{\rm ren}\over m_Q} \,,\qquad
   (\Delta C_2)_{\rm ren} = {2\over 3}\,
   {(\Delta m_Q)_{\rm ren}\over m_Q}
\end{equation}
derived in Ref.~\cite{Chris}. The behaviour of the distribution
functions for large values of $\tau$ is $\widehat w_1(\tau)\sim
1/\tau$ and $\widehat w_2(\tau)\sim 1/\tau^2$. The slow fall-off of
$\widehat w_1(\tau)$ leads to a logarithmic divergence in
$C_1(m_Q^2)$, which must be removed by renormalization.

\begin{table}[t]
\centerline{\parbox{15cm}{\caption{\label{tab:2}
Parameters related to the distribution function $\widehat w_2(\tau)$,
which is relevant for the ratio of heavy meson decay constants.}}}
\vspace{0.5cm}
\centerline{\begin{tabular}{l|cccccc}
\hline\hline
\rule[-0.2cm]{0cm}{0.7cm} & $\mu_{\rm BLM}^V$ &
 $\mu_{\rm BLM}^{\overline{\rm MS}}$ & $\alpha_s(\mu_{\rm BLM}^2)$
 & $\sigma_\tau$ & $\Delta$ & $\delta_{\rm BLM}$ \\
\hline
\rule[-0.1cm]{0cm}{0.6cm} $f_B^{\rm stat}/f_{B^*}^{\rm stat}$ &
 $0.472\,m_b$ & $0.205\,m_b$ & 0.41 & 2.798 & 7.830 & 0.67 \\
\hline\hline
\end{tabular}}
\vspace{0.5cm}
\end{table}

As an application of these results, consider the ratio of the decay
constants of the pseudoscalar and vector mesons $B$ and $B^*$ in the
so-called static limit, where terms of order $\Lambda_{\rm QCD}/m_b$
are neglected on the level of hadronic matrix elements. In this
limit, one finds \cite{subl}
\begin{equation}
   {f_B^{\rm stat}\over f_{B^*}^{\rm stat}}
   = 1 + {C_2(m_Q^2)\over C_1(m_Q^2)}
   = 1 + \int\limits_0^\infty\!{\rm d}\tau\,
   \widehat w_2(\tau)\,{\alpha_s(\tau e^C m_b^2)\over 4\pi}
   + \dots \,,
\end{equation}
where the ellipses represent terms not resummed in our approach.
Since $C_1(m_Q^2)=1+O(1/\beta_0)$, the distribution function is the
same as for the matching coefficient $C_2(m_Q^2)$.

\begin{figure}[htb]
   \vspace{0.5cm}
   \epsfysize=6cm
   \centerline{\epsffile{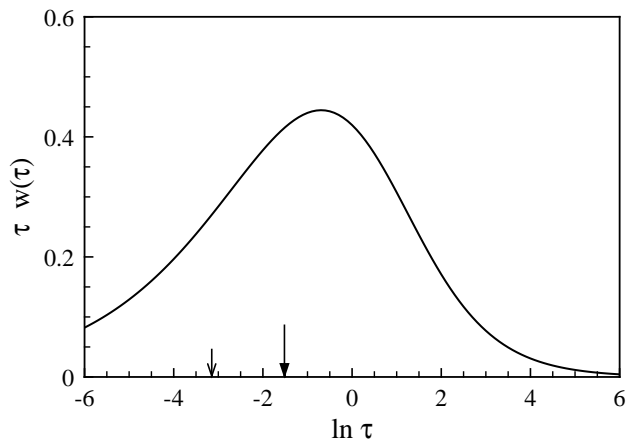}}
   \centerline{\parbox{13cm}{\caption{\label{fig:2}
Distribution function for the ratio $f_B^{\rm stat}/f_{B^*}^{\rm
stat}$. The right arrow indicates the BLM scale, the left one the
factorization point.}}}
\end{figure}

For the distribution function $\widehat w_2(\tau)$ we compute
\begin{eqnarray}
   N &=& {C_F\over 2} = {2\over 3} \,, \nonumber\\
   \langle \ln\tau \rangle &=& - {3\over 2} \,, \nonumber\\
   \langle \ln^2\!\tau \rangle &=& {7\over 2} + {2\pi^2\over 3} \,.
\end{eqnarray}
The resulting value of the BLM scale and some other parameters are
summarized in Table~\ref{tab:2}. Since in this case the distribution
function is positive definite, the parameter $\sigma_\tau =
\sqrt{\Delta}$ corresponds to the width of the distribution function.
The width is very large, about three units in $\ln\tau$. This is
clearly reflected in the shape of the function $\tau\,\widehat
w_2(\tau)$ shown in Fig.~\ref{fig:2}. As indicated by the position of
the arrows, the BLM scale is rather low, close to the factorization
point; the distribution is broad and extends well into the infrared
region. As a consequence, the convergence of the series is bad, as
reflected in the large value of the parameter $\delta_{\rm BLM}$ in
Table~\ref{tab:2}.

In the case of the ratio $m_Q/m_Q^{\rm R}$ considered above, the BLM
scale was low since the high-momentum contributions were removed, in
an ad hoc way, by the renormalization of the bare quark mass. On the
contrary, in the present case there is a physical reason why the
``typical'' momenta in the one-loop calculation are much lower than
the ``natural'' scale $m_b$. The flavour-changing vector and axial
vector currents are partially conserved, implying that they do not
receive radiative corrections from scales much above the masses of
their component fields \cite{Prep}. Roughly speaking, then, the
matching coefficients receive contributions from scales $0<\mu<m_Q$,
which is in fact the behaviour reflected in Fig.~\ref{fig:2}. (Recall
that $\mu=m_Q$ corresponds to $\ln\tau=0$ in the V scheme, and
$\ln\tau=5/3$ in the $\overline{\rm MS}$ scheme.) How fast the
distribution falls off in the infrared region is determined by the
location of the nearest infrared renormalon, which determines the
low-$\tau$ behaviour of the distribution function. Since in the
present case the nearest renormalon is located at $u=1/2$ in the
Borel plane, i.e.\ at the smallest possible value of $u$, the
fall-off is the slowest possible one, and thus there are substantial
contributions from the low-momentum region.

\subsection{Matching coefficients of heavy--heavy currents}
\label{sec:3.3}

Let us now consider a two-scale problem, namely the matching of
flavour-changing currents containing two heavy quarks onto their
counterparts in the heavy quark effective theory. At the so-called
zero recoil point, where the two heavy quarks move at the same
velocity, one finds \cite{review}
\begin{eqnarray}
   \bar c\,\gamma^\mu b &\to& \eta_V\,\bar h_v^c\gamma^\mu h_v^b
    + O(1/m_Q^2) \,, \nonumber\\
   \phantom{ \bigg[ }
   \bar c\,\gamma^\mu\gamma_5\,b &\to& \eta_A\,\bar h_v^c\gamma^\mu
    \gamma_5\,h_v^b + O(1/m_Q^2) \,.
\end{eqnarray}
The coefficients $\eta_V$ and $\eta_A$ take into account finite
renormalization effects. The coefficient $\eta_A$ of the axial vector
current plays a crucial role in the model-independent determination
of $|\,V_{cb}|$ from $B\to D^*\ell\,\bar\nu_\ell$ decays
\cite{Vcbnew}.

We choose $M=\sqrt{m_b m_c}$ as the ``natural'' mass scale and define
the ratio $z=m_c/m_b$. It is convenient to study the quantities
\begin{equation}
   S_V(m_b m_c,z) = \eta_V - 1 \,,\qquad
   S_{V-A}(m_b m_c,z) = \eta_V - \eta_A \,,
\end{equation}
since the structure of the perturbative corrections is very similar
for $\eta_V$ and $\eta_A$. The corresponding Borel transforms in the
large-$\beta_0$ limit are \cite{Chris,etaVA}
\begin{eqnarray}\label{SVSA}
   \widehat S_V(u,z) &=& C_F\,
    {\Gamma(u)\,\Gamma(1-2u)\over\Gamma(2-u)}\,\Bigg\{
    {2(1+u)\over 2-u}\,{z^u - z^{1-u}\over 1-z}
    + {2(1-u)\over 1+2u}\,{z^{-u} - z^{1+u}\over 1-z} \nonumber\\
   &&\mbox{}- {1+z\over 1-z}\,\big( z^u - z^{-u} \big)
    - {3(1-u^2)\over 2-u}\,\big( z^u + z^{-u} \big) \Bigg\} \,,
    \nonumber\\
   \widehat S_{V-A}(u,z) &=& 4 C_F\,
    {\Gamma(1+u)\,\Gamma(1-2u)\over\Gamma(3-u)}\,
    {z^u - z^{1-u}\over 1-z} \,.
\end{eqnarray}
After a straightforward calculation, we obtain the distribution
functions
\begin{eqnarray}
   \widehat w_V(\tau,z) &=& C_F\,\Bigg\{
    - {1-\tau\over 2}\,{(1-z)^2\over z}
    - {3\over 4}\,\tau\,{(1-z^2)^2\over z^2} \nonumber\\
   &&\quad\mbox{}- {3\over 4}\,{z\over\sqrt{1+4/(\tau z)}}
    - {3\over 4}\,{1\over z\sqrt{1+4 z/\tau}} \nonumber\\
   &&\quad\mbox{}+ \sqrt{1+{4\over\tau z}}\,\bigg[
    - {z\over 4} + {3\over 4}\,\tau\,z^2
    - {2\over\tau}\,{z\over 1-z} + {\tau\over 2}\,{z^2\over 1-z}
    \bigg] \nonumber\\
   &&\quad\mbox{}+ \sqrt{1+{4 z\over\tau}}\,\bigg[
    - {1\over 4 z} + {3\over 4}\,{\tau\over z^2}
    + {2\over\tau}\,{1\over 1-z} - {\tau\over 2}\,{1\over z(1-z)}
    \bigg] \Bigg\} \,, \nonumber\\
   && \nonumber\\
   \widehat w_{V-A}(\tau,z) &=& 2 C_F\,\Bigg\{
    1 + \tau\,{1+z+z^2\over 3 z} \nonumber\\
    &&\quad\mbox{}- {1\over 3(1-z)}\,\Bigg[
    \bigg( 1 + {\tau\over 2} \bigg)\,\sqrt{1+{4 z\over\tau}}
    - z (1+\tau z)\,\sqrt{1+{4\over\tau z}} \Bigg] \Bigg\} \,.
\end{eqnarray}
The asymptotic behaviour for small values of $\tau$ is
\begin{eqnarray}
   \widehat w_V(\tau,z) &=& - C_F\,{(1-z)^2\over 2 z}
    + O(\sqrt{\tau}) \,, \nonumber\\
   \phantom{ \bigg[ }
   \widehat w_{V-A}(\tau,z) &=& 2 C_F + O(\sqrt{\tau}) \,.
\end{eqnarray}
It is associated with infrared renormalon poles at $u=1$ in the Borel
plane. In fact, a careful investigation of (\ref{SVSA}) shows that
there are no poles at $u=1/2$. The corresponding renormalon
ambiguities in the matching coefficients are
\begin{eqnarray}
   (\Delta\eta_V)_{\rm ren} &=& - {2\over 3\beta_0}\,\Bigg(
    {\Lambda_V\over m_c} - {\Lambda_V\over m_b} \Bigg)^2 \,,
    \nonumber\\
   (\Delta\eta_A)_{\rm ren} &=& - {2\over 3\beta_0}\,\Bigg(
    {\Lambda_V\over m_c} + {\Lambda_V\over m_b} \Bigg)^2 \,.
\end{eqnarray}
This agrees with the results obtained in Ref.~\cite{Chris}. For large
values of $\tau$, both distribution functions fall off proportional
to $1/\tau^2$.

\begin{table}[t]
\centerline{\parbox{15cm}{\caption{\label{tab:3}
Parameters obtained from the distribution functions for the matching
coefficients $\eta_V$ and $(\eta_V-\eta_A)$ of heavy quark currents.
We use $m_b=4.8$~GeV and $m_c=1.44$~GeV.}}}
\vspace{0.5cm}
\centerline{\begin{tabular}{l|cccccc}
\hline\hline
\rule[-0.2cm]{0cm}{0.7cm} & $\mu_{\rm BLM}^V$ &
 $\mu_{\rm BLM}^{\overline{\rm MS}}$ & $\alpha_s(\mu_{\rm BLM}^2)$
 & $\sigma_\tau$ & $\Delta$ & $\delta_{\rm BLM}$ \\
\hline
\rule[-0.1cm]{0cm}{0.6cm} $\eta_V$ & $2.117\sqrt{m_b m_c}$ &
 $0.920\sqrt{m_b m_c}$ & 0.27 & 1.059 & 1.121 & 0.04 \\
\rule[-0.1cm]{0cm}{0.6cm} $\eta_V-\eta_A$ & $1.445\sqrt{m_b m_c}$ &
 $0.628\sqrt{m_b m_c}$ & 0.31 & 2.069 & 4.281 & 0.19 \\
\hline\hline
\end{tabular}}
\vspace{0.5cm}
\end{table}

\begin{figure}[htb]
   \vspace{0.5cm}
   \epsfysize=6cm
   \centerline{\epsffile{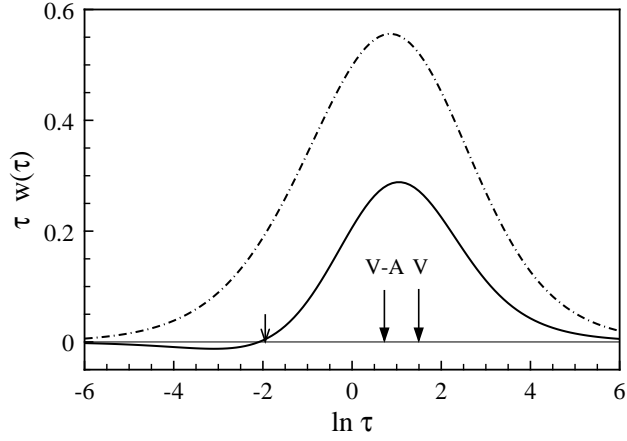}}
   \centerline{\parbox{13cm}{\caption{\label{fig:3}
Distribution functions for $\eta_V$ (solid line) and
$(\eta_V-\eta_A)$ (dashed-dotted line). The long arrows show the BLM
scales, the short arrow indicates the factorization point.}}}
\end{figure}

With the above distribution functions, we compute
\begin{eqnarray}
   N &=& {C_F\over 4}\,\phi(z) \simeq 0.236 \,, \nonumber\\
   \langle \ln\tau \rangle &=& {3\over 2} \,, \nonumber\\
   \langle \ln^2\!\tau \rangle &=& {2\pi^2\over 3} + {9\over 2}
    + {1\over 3}\,\ln^2\!z - 4\,{\ln^2\!z\over\phi(z)}
    \simeq 3.372 \,,
\end{eqnarray}
for $\widehat w_V(\tau,z)$, and
\begin{eqnarray}
   N &=& {C_F\over 2} = {2\over 3} \,, \nonumber\\
   \langle \ln\tau \rangle &=& {1\over 2} + {1\over 3}\,\phi(z)
    \simeq 0.736 \,, \nonumber\\
   \langle \ln^2\!\tau \rangle &=& {2\pi^2\over 3} - {5\over 2}
     + \ln^2\!z - \phi(z) \simeq 4.821 \,,
\end{eqnarray}
for $\widehat w_{V-A}(\tau,z)$, where
\begin{equation}
   \phi(z) = - 3\,{1+z\over 1-z}\,\ln z - 6
   = {\ln^2\!z\over 2} - {\ln^4\!z\over 120} + O(\ln^6\!z) \,.
\end{equation}
We have used the value $z=m_c/m_b=0.3$ in the numerical analysis. The
corresponding BLM scales and the values of the parameters
$\sigma_\tau$, $\Delta$ and $\delta_{\rm BLM}$ are given in
Table~\ref{tab:3}. The distribution functions are shown in
Fig.~\ref{fig:3}. In both cases the BLM scales are comfortably large
and are clearly separated from the factorization point, which
corresponds to $\tau=\lambda^2/m_b m_c$ with $\lambda=1$ GeV. We note
that the scales $\mu=m_b$ and $\mu=m_c$ correspond to $\ln\tau=\pm
1.204-C$, with $C=0$ in the V scheme and $C=-5/3$ in the
$\overline{\rm MS}$ scheme. The distributions fall off rapidly in the
infrared region. Therefore, both series converge much better than in
the case of the heavy--light current considered in the previous
section.

The physical reason for this behaviour is again related to current
conservation. As before, the currents are conserved in the
ultraviolet region, i.e.\ for scales $\mu\gg m_b$. But in contrast to
heavy--light currents, currents containing two heavy quarks moving at
the same velocity are also conserved in the infrared region, i.e.\
for $\mu\ll m_c$. In fact, at zero recoil the anomalous dimension
associated with such currents in the heavy quark effective theory
vanishes to all orders in perturbation theory \cite{Korc,Falk}. This
means that the matching coefficients receive sizeable contributions
only from scales $m_c<\mu<m_b$, which is in accordance with the
behaviour exhibited in Fig.~\ref{fig:3}.

\subsection{Numerical analysis}
\label{sec:3.4}

Let us now analyse our results. For each of the quantities $S(M^2)$
considered in the previous sections, we compare the following
approximations: the one- and (partial) two-loop expressions evaluated
using the ``natural'' scale $M$ in the running coupling constant, the
one-loop expression evaluated using the BLM scale, the truncated
series including the first correction to the BLM scheme given by the
term proportional to $\Delta$ in (\ref{Sappr}), the partial
resummation provided by the integral over the distribution function.
In the latter case we use the principle value prescription to
regularize the Landau pole in the running coupling constant. We
define
\begin{eqnarray}\label{apprS}
   S_{\rm 1-loop}(M^2) &=& N\,{\alpha_s(M^2)\over\pi} \,,
    \nonumber\\
   S_{\rm 2-loop}(M^2) &=& N\,{\alpha_s(M^2)\over\pi}\,\Bigg\{
    1 - \Big( C + \langle \ln\tau \rangle \Big)\,
    {\beta_0\,\alpha_s(M^2)\over 4\pi} \Bigg\} \,, \nonumber\\
   S_{\rm BLM}(M^2) &=& N\,{\alpha_s(\mu_{\rm BLM}^2)\over\pi} \,,
    \nonumber\\
   S_{\rm BLM^*}(M^2) &=& N\,{\alpha_s(\mu_{\rm BLM}^2)\over\pi}\,
    \Bigg\{ 1 + \Delta\,\Bigg( {\beta_0\,\alpha_s(\mu_{\rm BLM}^2)
    \over 4\pi} \Bigg)^2 \Bigg\} \,, \nonumber\\
   S_{\rm res}(M^2) &=& N\,\Bigg\langle
    {\alpha_s(\tau e^C M^2)\over\pi} \Bigg\rangle \,.
\end{eqnarray}
Note that $S_{\rm 2-loop}$ takes into account only part of the
two-loop corrections, namely those proportional to
$\beta_0\,\alpha_s^2$. As pointed out in the introduction, for the
quantities considered here it is known that the remaining two-loop
corrections are very small. We use the one-loop expression for the
running coupling constant in the $\overline{\rm MS}$ scheme with
$\Lambda_5=111$ MeV, $\Lambda_4=150$~MeV and $\Lambda_3=177$ MeV, so
that the coupling constant is continuous when one crosses the quark
thresholds at $m_b=4.8$ GeV and $m_c=1.44$ GeV. For reference
purposes we quote that $\alpha_s(m_b^2)\simeq 0.218$ and
$\alpha_s(m_c^2)\simeq 0.333$. Our results are shown in
Table~\ref{tab:5}. In the last column we give the value of the
renormalon ambiguity defined in (\ref{Drenorm}). To obtain it we use
$\beta_0=9$ and $\Lambda_V=408$ MeV, corresponding to $n_f=3$. The
value of $\Delta S_{\rm ren}$ should be considered as an estimate of
the intrinsic ambiguity in the result for the resummed series~$S_{\rm
res}$.

\begin{table}[t]
\centerline{\parbox{15cm}{\caption{\label{tab:5}
Comparison of various approximations for the quantities shown in the
first column.}}}
\vspace{0.5cm}
\centerline{\begin{tabular}{l|cccccc}
\hline\hline
\rule[-0.2cm]{0cm}{0.7cm} & $S_{\rm 1-loop}$ & $S_{\rm 2-loop}$ &
 $S_{\rm BLM}$ & $S_{\rm BLM^*}$ & $S_{\rm res}$ &
 $\Delta S_{\rm ren}$ \\
\hline
\rule[-0.1cm]{0cm}{0.6cm} $m_b/m_b^{\rm R1}$ & 0.092 & 0.146 &
 0.234 & 0.259 & 0.178 & 0.025 \\
\rule[-0.1cm]{0cm}{0.6cm} $m_b/m_b^{\rm R2}$ & 0.219 & 0.271 &
 0.288 & 0.336 & 0.304 & 0.025 \\
\rule[-0.1cm]{0cm}{0.6cm} $m_b/\overline{m}_b(m_b)$ & 0.092 &
 0.155 & 0.310 & $-0.059\phantom{-}$ & 0.188 & 0.025 \\
\rule[-0.1cm]{0cm}{0.6cm} $f_B^{\rm stat}/f_{B^*}^{\rm stat}$ &
 0.046 & 0.067 & 0.086 & 0.144 & 0.076 & 0.017 \\
\rule[-0.1cm]{0cm}{0.6cm} $\eta_V$ & 0.020 & 0.020 & 0.020 &
 0.021 & 0.023 & $-0.003\phantom{-}$ \\
\rule[-0.1cm]{0cm}{0.6cm} $\eta_V-\eta_A$ & 0.056 & 0.065 &
 0.067 & 0.079 & 0.081 & 0.007 \\
\hline\hline
\end{tabular}}
\vspace{0.5cm}
\end{table}

We observe that for the ratios $m_b/\overline{m}_b(m_b)$ and
$f_B^{\rm stat}/f_{B^*}^{\rm stat}$, where the BLM prescription gives
very low scales and the corrections to the BLM scheme are large, the
resummation leads to results similar to the two-loop approximation.
Thus, higher-order corrections are smaller than indicated by the BLM
prescription. Nevertheless, in these cases the ambiguity due to the
presence of the nearby infrared renormalon at $u=1/2$ is quite
significant. For the matching constants $\eta_V$ and $\eta_A$ the
resummation follows the tendency indicated by the BLM scheme, and the
results are close to what one obtains taking into account the leading
correction to that scheme. The renormalon ambiguities are smaller in
this case, since the nearest infrared renormalon pole is located at
$u=1$.

\begin{table}[t]
\centerline{\parbox{15cm}{\caption{\label{tab:6}
Comparison of the BLM scale with the scale $\mu_*$ corresponding to
the partial resummation of the series. All values refer to the
$\overline{\rm MS}$ scheme. In the V scheme, the scales are larger by
a factor $e^{5/6}\simeq 2.3$}}}
\vspace{0.5cm}
\centerline{\begin{tabular}{l|cccc}
\hline\hline
\rule[-0.2cm]{0cm}{0.7cm} & $M$ & $\mu_{\rm BLM}/M$ & $\mu_*/M$ &
 $\mu_*/\mu_{\rm BLM}$ \\
\hline
\rule[-0.1cm]{0cm}{0.6cm} $m_b/m_b^{\rm R1}$ & $m_b$ & 0.131 &
 0.195 & 1.49 \\
\rule[-0.1cm]{0cm}{0.6cm} $m_b/m_b^{\rm R2}$ & $m_b$ & 0.435 &
 0.378 & 0.87 \\
\rule[-0.1cm]{0cm}{0.6cm} $m_b/\overline{m}_b(m_b)$ & $m_b$ &
 0.096 & 0.179 & 1.86 \\
\rule[-0.1cm]{0cm}{0.6cm} $f_B^{\rm stat}/f_{B^*}^{\rm stat}$ &
 $m_b$ & 0.205 & 0.259 & 1.26 \\
\rule[-0.1cm]{0cm}{0.6cm} $\eta_V$ & $\sqrt{m_b m_c}$ & 0.920 &
 0.690 & 0.75 \\
\rule[-0.1cm]{0cm}{0.6cm} $\eta_V-\eta_A$ & $\sqrt{m_b m_c}$ &
 0.628 & 0.418 & 0.67 \\
\hline\hline
\end{tabular}}
\vspace{0.5cm}
\end{table}

Another way of comparing our resummation to the BLM scheme is to
define, for each series, a scale $\mu_*$ such that the one-loop
correction evaluated using that scale reproduces the resummed series.
Hence, we write
\begin{eqnarray}
   {m_b\over m_b^{\rm R}} &=& 1 + \bigg( {1\over 2} + r \bigg)\,
    {\alpha_s(\mu_*)\over\pi} \,, \nonumber\\
   {f_B^{\rm stat}\over f_{B^*}^{\rm stat}} &=& 1
    + {2\over 3}\,{\alpha_s(\mu_*)\over\pi} \,, \nonumber\\
   \eta_V &=& 1 + {\phi(m_c/m_b)\over 3}\,
    {\alpha_s(\mu_*)\over\pi} \,, \nonumber\\
   \eta_V - \eta_A &=& {2\over 3}\,{\alpha_s(\mu_*)\over\pi} \,.
\end{eqnarray}
In Table~\ref{tab:6} these scales are compared to the BLM
scales. Note that in the cases with the lowest BLM scale, namely for
the ratios of masses and decay constants, the resummation leads to a
larger scale $\mu_*>\mu_{\rm BLM}$.

\begin{table}[t]
\centerline{\parbox{15cm}{\caption{\label{tab:5a}
Comparison of resummed ``perturbative'' series $S_{\rm res}$ with a
model calculation of the full series including long-distance effects,
$S_{\rm tot}=S_{\rm sd}(\lambda) + S_{\rm ld}(\lambda)$. The notation
is such that the main values given for $S_{\rm tot}$, $S_{\rm ld}$
and $\Lambda$ correspond to $\alpha_s(0)=2$, whereas the corrections
indicated as super- and subscripts refer to $\alpha_s(0)=1$ and 4,
respectively. We use $\lambda=1$ GeV for the factorization scale.}}}
\vspace{0.5cm}
\centerline{\begin{tabular}{l|cc|ccc}
\hline\hline
\rule[-0.2cm]{0cm}{0.7cm} & $S_{\rm res}$ & $S_{\rm tot}$ &
 $S_{\rm sd}(\lambda)$ & $S_{\rm ld}(\lambda)$ &
 $\Lambda(\lambda)$ [MeV] \\
\hline
\rule[-0.1cm]{0cm}{0.6cm} $m_b/m_b^{\rm R1}$ & 0.178 &
 $0.243^{-0.039}_{+0.046}$ & 0.138 & $0.105^{-0.039}_{+0.046}$ &
 $501^{-186}_{+221}$ \\
\rule[-0.1cm]{0cm}{0.6cm} $m_b/m_b^{\rm R2}$ & 0.304 &
 $0.369^{-0.039}_{+0.046}$ & 0.264 & $0.105^{-0.039}_{+0.046}$ &
 $501^{-186}_{+221}$ \\
\rule[-0.1cm]{0cm}{0.6cm} $m_b/\overline{m}_b(m_b)$ & 0.188 &
 $0.252^{-0.039}_{+0.046}$ & 0.147 & $0.105^{-0.039}_{+0.046}$ &
 $501^{-186}_{+221}$ \\
\rule[-0.1cm]{0cm}{0.6cm} $f_B^{\rm stat}/f_{B^*}^{\rm stat}$ &
 0.076 & $0.119^{-0.023}_{+0.029}$& 0.057 &
 $0.062^{-0.023}_{+0.029}$ & $300^{-111}_{+138}$ \\
\rule[-0.1cm]{0cm}{0.6cm} $\eta_V$ & 0.023 &
 $0.020^{+0.001}_{-0.001}$ & 0.024 & $-0.004^{+0.001}_{-0.001}
 \phantom{-}$ & $154^{-21}_{+25}$ \\
\rule[-0.1cm]{0cm}{0.6cm} $\eta_V-\eta_A$ & 0.081 &
 $0.083^{-0.006}_{+0.005}$ & 0.064 & $0.019^{-0.006}_{+0.005}$ &
 $363^{-67}_{+47}$ \\
\hline\hline
\end{tabular}}
\vspace{0.5cm}
\end{table}

Our next goal is to obtain an estimate for the relative and absolute
size of the short- and long-distance contributions to the various
quantities considered above. To this end we introduce a factorization
scale $\lambda=1$ GeV and evaluate separately the two integrals
$S_{\rm sd}$ and $S_{\rm ld}$ defined in (\ref{separ}). The
factorization scale is chosen such that the value of the coupling
constant $\alpha_s(e^C\lambda^2)/\pi\simeq 0.25$ is still in the
perturbative regime (note that $e^{C/2}\lambda\simeq 0.43$ GeV in the
$\overline{\rm MS}$ scheme). In order to model the long-distance
contribution, we guess a ``realistic'' behaviour of the coupling
constant in the infrared region. We use a modified version of the
running coupling constant, which exhibits freezing for $\mu\to 0$:
\begin{equation}
   \alpha_s(e^C\mu^2) = {4\pi\over\beta_0\,
   \ln(c+\mu^2/\Lambda_V^2)} \,.
\end{equation}
We shall investigate the cases where $c$ is adjusted so that
$\alpha_s(0)=1$, 2 and 4 (in the $\overline{\rm MS}$ scheme), and
interpret the dependence of the results on $c$ as a measure of the
model dependence. In Table~\ref{tab:5a} we compare the sum $S_{\rm
tot}=S_{\rm sd}+S_{\rm ld}$ to the ``perturbative'' resummation
obtained using the one-loop running coupling constant regulated with
a principle value prescription. We also give the results for the
short- and long-distance contributions separately. For each case, we
write the long-distance contribution as a power correction,
$|\,S_{\rm ld}|=(\Lambda/M)^{2k}$, and quote the value of the
low-energy scale $\Lambda$. We observe that the long-distance
contribution is large, as big as the short-distance one, for the
ratios of the heavy quark masses (in the schemes R1 and
$\overline{\rm MS}$) and for the ratio of the decay constants. The
reason is that the nearest infrared renormalon is located at $u=1/2$,
leading to non-perturbative corrections suppressed by only one power
of the large mass scale $m_b$. In general, the scales $\Lambda$
associated with the long-distance contributions are typical
low-energy scales of QCD.

\begin{figure}[htb]
   \vspace{0.5cm}
   \epsfysize=6cm
   \centerline{\epsffile{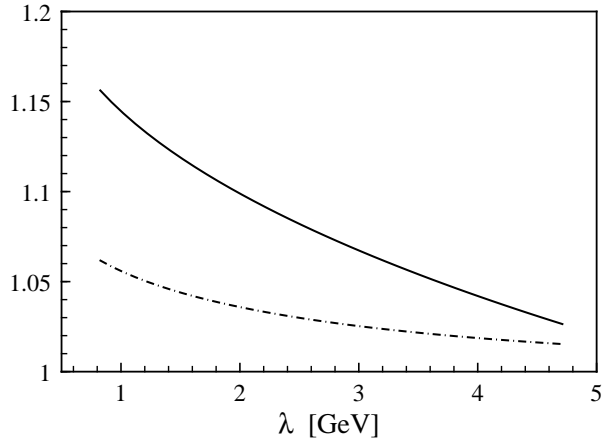}}
   \centerline{\parbox{13cm}{\caption{\label{fig:lam1}
Short-distance contributions to $m_b(\lambda)/\overline{m}_b(m_b)$
(solid line) and $(f_B/f_{B^*})(\lambda)$ (dashed-dotted line) as a
function of the factorization scale.}}}
\end{figure}

\begin{figure}[htb]
   \vspace{0.5cm}
   \epsfysize=6cm
   \centerline{\epsffile{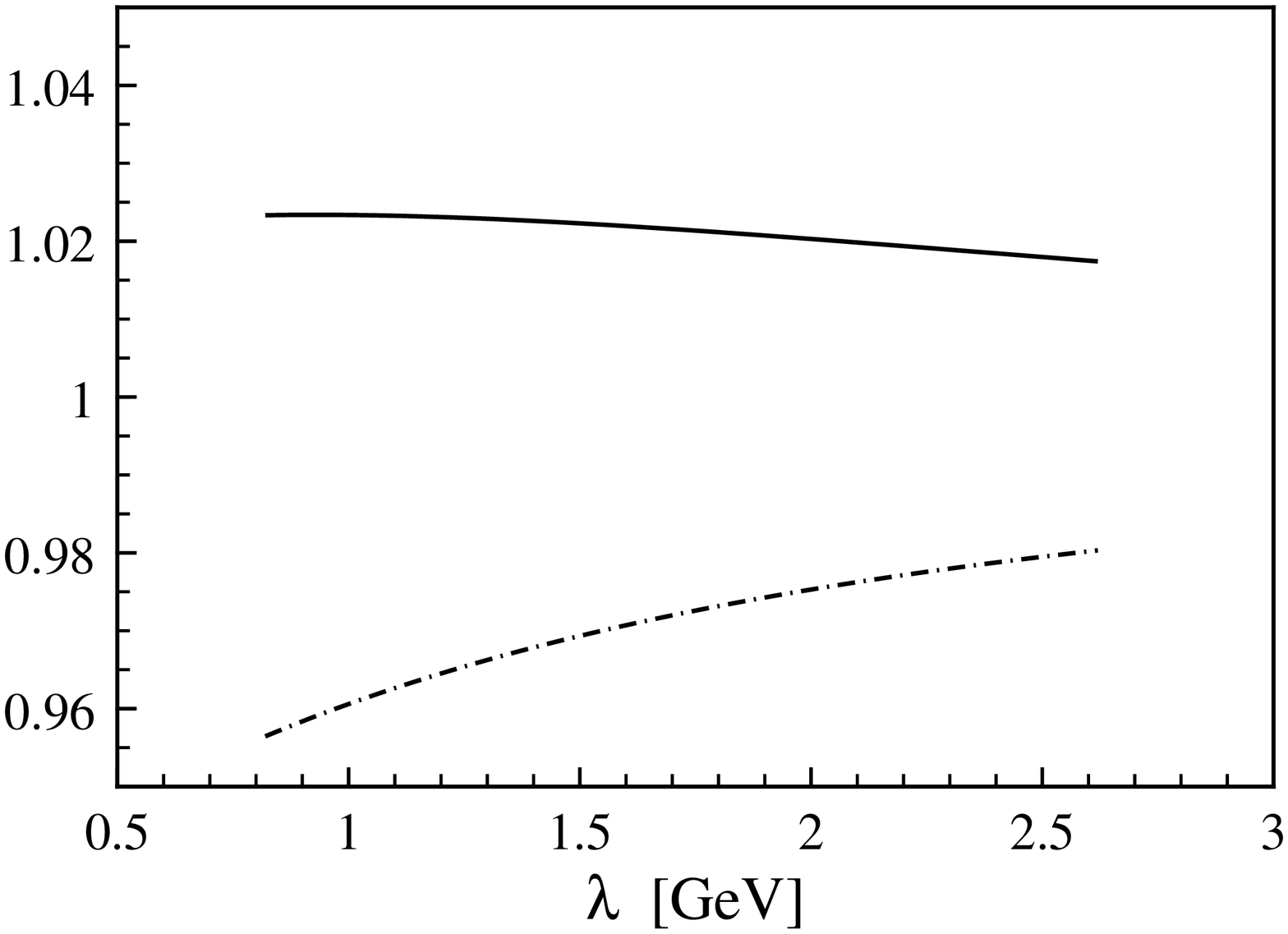}}
   \centerline{\parbox{13cm}{\caption{\label{fig:lam2}
Short-distance contributions to $\eta_V(\lambda)$ (solid line) and
$\eta_A(\lambda)$ (dashed-dotted line) as a function of the
factorization scale.}}}
\end{figure}

In Figs.~\ref{fig:lam1} and \ref{fig:lam2} we show our predictions
for the short-distance contributions to the quantities
$m_b/\overline{m}_b(m_b)$, $f_B/f_{B^*}$, $\eta_V$ and $\eta_A$ as a
function of the factorization scale in the range
$\lambda_0<\lambda<M$, where $\lambda_0\simeq 0.823$ GeV is the point
where $\alpha_s(e^C\lambda_0^2)=1$. These results will become useful
once non-perturbative calculations of matrix elements performed with
a hard ultraviolet cutoff become available. The idea to introduce a
hard factorization scale to organize the heavy quark expansion has
been put forward recently by Bigi et al.\ \cite{Bigiren}. Our
approach provides a consistent framework to implement this proposal.
Let us illustrate this with two examples of phenomenological
importance. The first is the ratio of the physical meson decay
constants $f_B$ and $f_{B^*}$, which is related to the ratio defined
in the static limit by
\begin{equation}\label{fexp}
   {f_B\over f_{B^*}} = {f_B^{\rm stat}\over f_{B^*}^{\rm stat}}
    + {A\over m_b} + O(1/m_b^2) \,.
\end{equation}
The non-perturbative parameter $A$ can be defined in terms of
hadronic matrix elements of dimension-four operators in the heavy
quark effective theory \cite{subl}, which have to be estimated using
non-perturbative techniques such as lattice gauge theory or QCD sum
rules. These matrix elements are linearly ultraviolet divergent. If
in the calculation of $A$ one introduces a hard ultraviolet cutoff
$\lambda$ in the same way as it was done for the perturbative
calculation in the static limit, these matrix elements contain
those long-distance contributions excluded in the short-distance
calculation. Hence, one obtains
\begin{equation}\label{fratio}
   {f_B\over f_{B^*}} = 1 + S_{\rm sd}(m_b^2,\lambda)
   + {A(\lambda)\over m_b} + O(1/m_b^2)
   \equiv {f_B\over f_{B^*}}(\lambda) + {A(\lambda)\over m_b}
   + O(1/m_b^2) \,.
\end{equation}
Our resummed expression for the short-distance contribution is
\begin{equation}
   S_{\rm sd}(m_b^2,\lambda) = \int\limits_{\lambda^2/m_b^2}^\infty\!
   {\rm d}\tau\,\widehat w_2(\tau)\,
   {\alpha_s(\tau e^C m_b^2)\over 4\pi} \,,
\end{equation}
where the distribution function $\widehat w_2(\tau)$ has been given
in (\ref{w1w2}). The $\lambda$-dependence in (\ref{fratio}) cancels
between the short- and long-distance pieces. Let us note that often
heavy quark expansions such as (\ref{fexp}) are written down using
dimensional regularization, in which case there is no clear
separation between short- and long-distance contributions. Then the
perturbative series contains an infrared renormalon at $u=1/2$, which
is exactly compensated by an ultraviolet renormalon in the parameter
$A$ \cite{Chris}. Again, the sum of all perturbative and
non-perturbative terms is unambiguous.

Our second example involves the matching factor $\eta_A$ for a
flavour-changing axial vector current containing two heavy quarks.
{}From the measurement of the recoil spectrum in the semi-leptonic
decay $B\to D^*\ell\,\bar\nu_\ell$ one can extract the product
$|\,V_{\rm cb}|\,{\cal F}(1)$, where ${\cal F}(1)$ denotes the value
of the hadronic form factor of the decay at the kinematic point of
zero recoil \cite{Vcbnew}. This form factor is usually factorized in
the form ${\cal F}(1)=\eta_A\,(1+\delta_{1/m^2})$, where the quantity
$\delta_{1/m^2}$ represents non-perturbative power corrections, which
from a conceptual point of view cannot be distinguished from the
long-distance contributions to $\eta_A$. Hence, to separate short-
and long-distance effects properly one should again introduce a
factorization scale, so that $\eta_A(\lambda)$ contains all
short-distance contributions, while $\delta_{1/m^2}(\lambda)$
accounts for long-distance effects. The problem is that so far all
calculations of $\delta_{1/m^2}$ were based on phenomenological
approaches that do not account for the $\lambda$-dependence
\cite{FaNe}--\cite{Shifs}. Therefore, we have to rely on a reasonable
guess for the factorization scale when we combine the most recent
estimate $\delta_{1/m^2}=-(5.5\pm 2.5)\%$ \cite{Vcbnew} with our
short-distance calculation. From Fig.~\ref{fig:lam2} we find that
$0.955<\eta_A(\lambda)<0.975$ for $0.8~\mbox{GeV} < \lambda <
2~\mbox{GeV}$, which we consider a conservative range of values for
the factorization scale. This yields
\begin{equation}
   {\cal F}(1) = \eta_A(\lambda)\,[ 1+\delta_{1/m^2}(\lambda) ]
   = 0.91\pm 0.03 \,,
\end{equation}
which is 2\% larger than the result obtained in Ref.~\cite{Shifs},
and 2\% smaller than the value quoted in Ref.~\cite{Vcbnew}. The
corresponding shift in the value of $|\,V_{\rm cb}|$ is at the level
of $10^{-3}$.

\section{Correlator of light vector currents}
\label{sec:4}

As an important example not related to heavy quarks, we investigate
the perturbative expansion of the correlator of two vector currents
in the euclidean region ($Q^2=-q^2>0$):
\begin{equation}
   i \int\!{\rm d}^4 x\,e^{i q\cdot x}\,\langle\,0\,|\,
   T\{ j^\mu(x),j^\nu(0) \}\,|\,0\,\rangle
   = (q^\mu q^\nu - q^2 g^{\mu\nu})\,\Pi(Q^2) \,,
\end{equation}
where $j^\mu=\bar q\,\gamma^\mu q$. For simplicity we shall consider
massless quarks. The momentum transfer $Q^2$ provides the large mass
scale. The derivative of $\Pi(Q^2)$ with respect to $Q^2$ is
ultraviolet convergent. As usual, we define the $D$-function
\begin{equation}\label{dQPi}
   D(Q^2) = 4\pi^2\,Q^2\,{{\rm d}\Pi(Q^2)\over{\rm d}Q^2}
   = 1 + S_{D}(Q^2) \,.
\end{equation}
In the large-$\beta_0$ limit, the Borel transform of the perturbative
series for $S_D(Q^2)$ is well known \cite{Bene,Broa}:
\begin{eqnarray}\label{SPi}
   \widehat S_D(u) &=& {32 C_F\over 2-u}\,
    \sum_{k=2}^\infty\,{(-1)^k\,k\over\big[ k^2-(1-u)^2\big]^2}
    \nonumber\\
   &=& 3 C_F\,\bigg\{ 1 + \bigg( {23\over 6} - 4\zeta(3) \bigg)\,u
    + \Big( 9 - 6\zeta(3) \Big)\,u^2 + O(u^3) \bigg\} \,.
\end{eqnarray}
To obtain the corresponding distribution function we start from the
first relation in (\ref{wSrel}) and set $u_0=1$. This gives
\begin{equation}
   \widehat w_D(\tau) = {8 C_F\over\pi}\,
   \sum_{k=2}^\infty\,(-1)^k\,{{\rm d}\over{\rm d}k}\,
   {1\over k^2-1}\,\int\limits_{-\infty}^\infty\!{\rm d}r\,
   e^{i r\ln\tau}\,\bigg( {1+i r\over r^2+k^2}
   - {1\over 1-i r} \bigg) \,.
\end{equation}
The integral can be performed closing the integration contour at
infinity; however, it is necessary to distinguish the cases
$\ln\tau>0$ and $\ln\tau<0$. We find
\begin{eqnarray}\label{wparpi}
   \widehat w_D(\tau) &=& 8 C_F\,\bigg\{
    \bigg( {7\over 4} - \ln\tau \bigg)\,\tau
    + (1+\tau)\,\Big[ L_2(-\tau) + \ln\tau\,\ln(1+\tau) \Big]
    \bigg\} \,;\quad \tau<1 \,, \nonumber\\
   && \nonumber\\
   \widehat w_D(\tau) &=& 8 C_F\,\bigg\{ 1 + \ln\tau
    + \bigg( {3\over 4} + {1\over 2}\,\ln\tau \bigg)\,{1\over\tau}
    \nonumber\\
   &&\qquad \mbox{}+ (1+\tau)\,\Big[ L_2(-\tau^{-1}) - \ln\tau\,
    \ln(1+\tau^{-1}) \Big] \bigg\} \,;\quad \tau>1 \,,
\end{eqnarray}
where $L_2(x)=-\int_0^x{{\rm d}y\over y}\ln(1-y)$ is the dilogarithm
function. The distribution function and its first three derivatives
are continuous at $\tau=1$, but higher derivatives are not. The
asymptotic behaviour for $\tau\to 0$ is given by
\begin{equation}\label{asy1}
   \widehat w_D(\tau) = 6 C_F\,\tau + O(\tau^2) \,,
\end{equation}
corresponding to the infrared renormalon pole at $u=2$ in the Borel
transform in (\ref{SPi}). The location of this renormalon is
consistent with the structure of the OPE for $\Pi(Q^2)$, in which
non-perturbative corrections appear first\footnote{The question
whether there is an infrared renormalon at $u=1$ in real QCD (beyond
the large-$\beta_0$ approximation) is, however, not completely
settled \protect\cite{Muel,doubt}.}
at order $1/Q^4$. This will be discussed in more detail below. For
large values of $\tau$ the distribution function behaves like
$\ln\tau/\tau^2$, so that the integral over the distribution function
is ultraviolet convergent.

\begin{figure}[htb]
   \vspace{0.5cm}
   \epsfysize=6cm
   \centerline{\epsffile{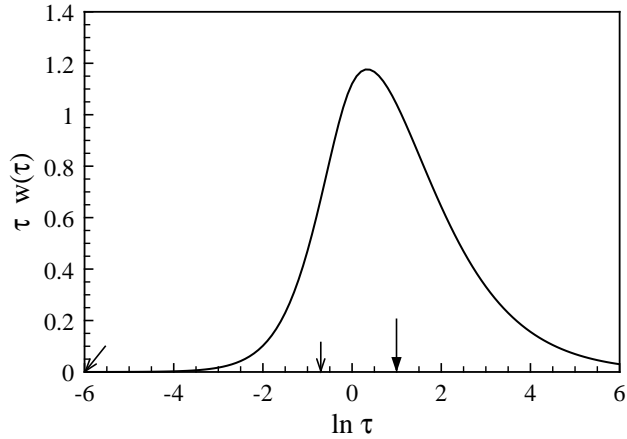}}
   \centerline{\parbox{13cm}{\caption{\label{fig:4}
Distribution function for the $D$-function. The long arrow shows the
BLM scale. The short arrows indicate the factorization point
$\tau=\lambda^2/Q^2$ for $Q_1^2=2~\mbox{GeV}^2$ (right) and
$Q_2^2=(20~\mbox{GeV})^2$ (left).}}}
\end{figure}

Using the above result, one can compute
\begin{eqnarray}\label{corints}
   N &=& {3\over 4}\,C_F = 1 \,, \nonumber\\
   \langle \ln\tau \rangle &=& 4\zeta(3) - {23\over 6}
    \simeq 0.975 \,, \nonumber\\
   \phantom{ \bigg[ }
   \langle \ln^2\!\tau \rangle &=& 18 - 12\zeta(3)\simeq 3.575 \,.
\end{eqnarray}
The resulting values for the BLM scale and for the parameters
$\sigma_\tau$, $\Delta$ and $\delta_{\rm BLM}$ are shown in
Table~\ref{tab:4} for both a small and a large value of $Q^2$. A
graphical representation of the distribution function is given in
Fig.~\ref{fig:4}.

\begin{table}[t]
\centerline{\parbox{15cm}{\caption{\label{tab:4}
Parameters obtained from the distribution function corresponding to
the $D$-function. We use $Q_1^2=2~\mbox{GeV}^2$ and
$Q_2^2=(20~\mbox{GeV})^2$.}}}
\vspace{0.5cm}
\centerline{\begin{tabular}{l|cccccc}
\hline\hline
\rule[-0.2cm]{0cm}{0.7cm} & $\mu_{\rm BLM}^V$ &
 $\mu_{\rm BLM}^{\overline{\rm MS}}$ & $\alpha_s(\mu_{\rm BLM}^2)$
 & $\sigma_\tau$ & $\Delta$ & $\delta_{\rm BLM}$ \\
\hline
\rule[-0.1cm]{0cm}{0.6cm} $D(Q_1^2)$ & $1.628\sqrt{Q_1^2}$ &
 $0.708\sqrt{Q_1^2}$ & 0.40 & 1.620 & 2.625 & 0.22 \\
\rule[-0.1cm]{0cm}{0.6cm} $D(Q_2^2)$ & $1.628\sqrt{Q_2^2}$ &
 $0.708\sqrt{Q_2^2}$ & 0.17 & 1.620 & 2.625 & 0.03 \\
\hline\hline
\end{tabular}}
\vspace{0.5cm}
\end{table}

For completeness we also discuss the resummation for the correlator
$\Pi(Q^2)$ itself. To obtain it, we integrate (\ref{dQPi}) and use
the fact that $S_D(Q^2)$ depends on $Q^2$ only through the running
coupling constant to find
\begin{equation}
   4\pi^2\,\Big[ \Pi(Q^2) - \Pi(Q_0^2) \Big]_{\rm res}
   = \ln{Q^2\over Q_0^2} + {1\over 2\pi}
   \int\limits_0^\infty\!{\rm d}\tau\,\widehat w_D(\tau)\,
   \int\limits_{\alpha_s(\tau e^C Q_0^2)}^{\alpha_s(\tau e^C Q^2)}\!
   {\rm d}\alpha_s\,{\alpha_s\over\beta(\alpha_s)} \,,
\end{equation}
where $\beta(\alpha_s) = {\rm d}\alpha_s(\mu^2)/{\rm d}\ln\mu$ is
the $\beta$-function, and $Q_0^2$ is some arbitrary reference scale,
which serves to subtract the ultraviolet divergence of $\Pi(Q^2)$. To
obtain the distribution function for $\Pi(Q^2)$ it is sufficient to
use the one-loop $\beta$-function,
\begin{equation}
   \beta(\alpha_s) = - {\beta_0\over 2\pi}\,\alpha_s^2 \,,
\end{equation}
which leads to
\begin{equation}\label{Pirep}
   4\pi^2\,\Big[ \Pi(Q^2) - \Pi(Q_0^2) \Big]_{\rm res}
   = \ln{Q^2\over Q_0^2} - {1\over\beta_0}\,\int\limits_0^\infty\!
   {\rm d}\tau\,\widehat w_D(\tau)\,
   \ln{\alpha_s(\tau e^C Q^2)\over\alpha_s(\tau e^C Q_0^2)} \,.
\end{equation}
Using an integration by parts we can bring this into the standard
form of the distribution function representation:
\begin{equation}
   4\pi^2\,\Big[ \Pi(Q^2) - \Pi(Q_0^2) \Big]_{\rm res}
   = \ln{Q^2\over Q_0^2}
   +\int\limits_0^\infty\!{\rm d}\tau\,\widehat w_\Pi(\tau)\,
   \Bigg\{ {\alpha_s(\tau e^C Q^2)\over 4\pi}
   - {\alpha_s(\tau e^C Q_0^2)\over 4\pi} \Bigg\} \,,
\end{equation}
with the distribution function
\begin{equation}
   \widehat w_\Pi(\tau) = - {1\over\tau}\,\int\limits_0^\tau\!
   {\rm d}\tau'\,\widehat w_D(\tau') \,.
\end{equation}
Performing the integral gives
\begin{eqnarray}\label{wPi}
   \widehat w_\Pi(\tau) &=& -4 C_F\,\bigg\{ 1 - \ln\tau
    + \bigg( {5\over 2} - {3\over 2}\ln\tau \bigg)\,\tau
    \nonumber\\
   &&\qquad\quad\mbox{}+ {(1+\tau)^2\over\tau}\,\Big[ L_2(-\tau)
    + \ln\tau\,\ln(1+\tau) \Big] \bigg\} \,;\quad \tau<1
    \,, \nonumber\\
   && \nonumber\\
   \widehat w_\Pi(\tau) &=& -4 C_F\,\bigg\{ 1 + \ln\tau
    + \bigg( {5\over 2} + {3\over 2}\ln\tau \bigg)\,{1\over\tau}
    \nonumber\\
   &&\qquad\quad\mbox{}+ {(1+\tau)^2\over\tau}\,\Big[
    L_2(-\tau^{-1}) - \ln\tau\,\ln(1+\tau^{-1}) \Big] \bigg\}
    \,;\quad \tau>1 \,.
\end{eqnarray}
The same result can also be obtained directly by starting from the
Borel transform of the correlator $\Pi(Q^2)$. The asymptotic
behaviour for small values of $\tau$ is
\begin{equation}
   \widehat w_\Pi(\tau) = -3 C_F\,\tau + O(\tau^2) \,,
\end{equation}
corresponding again to an infrared renormalon at $u=2$. For large
values of $\tau$ the distribution function falls off like $1/\tau$,
in accordance with the logarithmic ultraviolet divergence of the
unsubtracted correlator.

\begin{table}[t]
\centerline{\parbox{15cm}{\caption{\label{tab:D1}
Various approximations for the perturbative contribution to the
$D$-function. As previously $Q_1^2=2~\mbox{GeV}^2$ and
$Q_2^2=(20~\mbox{GeV})^2$.}}}
\vspace{0.5cm}
\centerline{\begin{tabular}{l|cccccc}
\hline\hline
\rule[-0.2cm]{0cm}{0.7cm} & $S_{\rm 1-loop}$ & $S_{\rm 2-loop}$ &
 $S_{\rm BLM}$ & $S_{\rm BLM^*}$ & $S_{\rm res}$ &
 $\Delta S_{\rm ren}$ \\
\hline
\rule[-0.1cm]{0cm}{0.6cm} $D(Q_1^2)$ & 0.107 & 0.125 & 0.128 &
 0.156 & 0.164 & 0.006 \\
\rule[-0.1cm]{0cm}{0.6cm} $D(Q_2^2)$ & 0.050 & 0.054 & 0.054 &
 0.055 & 0.055 & $1.5\times 10^{-7}$ \\
\hline\hline
\end{tabular}}
\vspace{0.5cm}
\end{table}

\begin{table}[t]
\centerline{\parbox{15cm}{\caption{\label{tab:D2}
Comparison of the BLM scale with the scale $\mu_*$ corresponding to
the full resummation of the series. All values refer to the
$\overline{\rm MS}$ scheme.}}}
\vspace{0.5cm}
\centerline{\begin{tabular}{l|cccc}
\hline\hline
\rule[-0.2cm]{0cm}{0.7cm} & $\mu_{\rm BLM}/\sqrt{Q^2}$ &
 $\mu_*/\sqrt{Q^2}$ & $\mu_*/\mu_{\rm BLM}$ \\
\hline
\rule[-0.1cm]{0cm}{0.6cm} $D(Q_1^2)$ & 0.708 & 0.485 & 0.685 \\
\rule[-0.1cm]{0cm}{0.6cm} $D(Q_2^2)$ & 0.708 & 0.621 & 0.878 \\
\hline\hline
\end{tabular}}
\vspace{0.5cm}
\end{table}

Let us now turn to the numerical analysis of our results. In
Table~\ref{tab:D1} we show the various approximations to the series
$S_D(Q^2)$, as defined in (\ref{apprS}). We also quote values for the
renormalon ambiguity
\begin{equation}
   \Delta D_{\rm ren} = (\Delta S_D)_{\rm ren}
   = {8\over\beta_0}\,\Bigg( {\Lambda_V^2\over Q^2} \Bigg)^2 \,.
\end{equation}
It is apparent that the effect of the resummation is more pronounced
in the case where $Q^2$ is low. In this case there are significant
corrections to the BLM scheme. For the resummed series $S_{\rm
res}(Q^2)$ we define a scale $\mu_*$ so that
\begin{equation}
   D(Q^2) = 1 + {\alpha_s(\mu_*^2)\over\pi} \,.
\end{equation}
This scale is compared to the BLM scale in Table~\ref{tab:D2}.
Finally, in Table~\ref{tab:D3} we evaluate the short- and
long-distance contributions to the $D$-function introducing a
factorization scale $\lambda=1$ GeV.

\begin{table}[t]
\centerline{\parbox{15cm}{\caption{\label{tab:D3}
Comparison of resummed ``perturbative'' series $S_{\rm res}$ with a
model calculation of the full series including long-distance effects,
$S_{\rm tot}=S_{\rm sd}(\lambda) + S_{\rm ld}(\lambda)$. We use
$\lambda=1$ GeV for the factorization scale.}}}
\vspace{0.5cm}
\centerline{\begin{tabular}{c|cc|ccc}
\hline\hline
\rule[-0.2cm]{0cm}{0.7cm} & $S_{\rm res}$ & $S_{\rm tot}$ &
 $S_{\rm sd}(\lambda)$ & $S_{\rm ld}(\lambda)$ &
 $\Lambda(\lambda)$ [MeV] \\
\hline
\rule[-0.1cm]{0cm}{0.6cm} $D(Q_1^2)$ & 0.164 &
 $0.146^{-0.009}_{+0.005}$ & 0.118 & $0.028^{-0.009}_{+0.005}$ &
 $581^{-56}_{+23}$ \\
\rule[-0.1cm]{0cm}{0.6cm} $D(Q_2^2)$ & 0.005 & 0.005 & 0.005 &
 $(1.6\pm 0.2)\times 10^{-6}$ & $713^{-31}_{+16}$ \\
\hline\hline
\end{tabular}}
\vspace{0.5cm}
\end{table}

{}From the asymptotic behaviour of the distribution function in
(\ref{asy1}) we conclude that the long-distance contribution to the
perturbative series scales like
\begin{equation}
   S_{\rm ld}(Q^2,\lambda) = {\Lambda^4(\lambda)\over(Q^2)^2} \,.
\end{equation}
It should be combined with non-perturbative contributions of the same
magnitude. For the euclidean correlator the OPE of the current
product $j^\mu(x)\,j^\nu(0)$ provides the framework for a systematic
incorporation of non-perturbative effects. At order $1/(Q^2)^2$ these
effects are parametrized by the gluon condensate \cite{SVZ}. Hence,
to this order we may write
\begin{eqnarray}
   D(Q^2) &=& 1 + S_{\rm sd}(Q^2,\lambda)
    + {\Lambda^4(\lambda)\over(Q^2)^2} + {2\pi\over 3(Q^2)^2}\,
    \langle\alpha_s\,G^2\rangle + \dots \nonumber\\
   &\equiv& 1 + S_{\rm sd}(Q^2,\lambda) + {2\pi\over 3(Q^2)^2}\,
    \langle\alpha_s\,G^2\rangle(\lambda) + \dots \,,
\end{eqnarray}
where the last equation defines the scale-dependent condensate
\begin{equation}\label{newGG}
   \langle\alpha_s\,G^2\rangle(\lambda)
   = \langle\alpha_s\,G^2\rangle
   + {3\over 2\pi}\,\Lambda^4(\lambda) \,.
\end{equation}
For $\lambda=1$ GeV, we find from Table~\ref{tab:D3} that the
``perturbative'' contribution to the gluon condensate is about
$0.1~\mbox{GeV}^4$, which is of the same order of magnitude as the
``genuine'' gluon condensate $\langle\alpha_s\,G^2\rangle$
\cite{SVZ}. In many practical applications of the OPE, and in
particular in the phenomenology of the QCD sum rules, it
is assumed that the ``perturbative'' contributions to the vacuum
condensates are much smaller than the ``genuine'' values of the
condensates and can be neglected \cite{SVZBuch}. Our result
(\ref{newGG}) provides a counter-example to this assertion.

\section{Conclusions}
\label{sec:6}

We have proposed an extension of the BLM scale-setting prescription,
which resums certain vacuum polarization insertions to all orders in
perturbation theory. Our approach is equivalent to performing
one-loop calculations with a running coupling constant, thereby
including much of the non-trivial asymptotic behaviour of a
perturbative series. The representation of the resummed series as an
integral over the running coupling constant with a distribution
function, as shown in (\ref{Sappr}), provides an intuitive picture of
the distribution of virtualities in a one-loop calculation. Much
insight can be gained from the knowledge of the distribution
function. Its behaviour for large and small values of the scale
parameter $\tau$ is related in a direct way to the ultraviolet and
infrared properties of the series. Moments of the distribution
function determine the size of higher-order coefficients.

By summing an infinite set of diagrams our scheme reaches beyond
perturbation theory. In particular, it provides a clear separation of
short- and long-distance effects. In any finite-order perturbative
calculation non-perturbative effects are implicitly present due to
low-momentum contributions in Feynman diagrams, but are not visible
as they are exponentially small in the coupling constant. Yet
perturbation theory ``knows'' about these contributions in the form
of infrared renormalon singularities, which make a perturbative
series non Borel summable. This means that attempts to resum the
series will lead to unavoidable ambiguities. In our scheme the
long-distance contributions can be explicitly separated, since it is
possible to introduce a hard momentum cutoff in a natural way. The
size of the long-distance contributions and their dependence on the
large mass scale of the problem is determined by the asymptotic
behaviour of the distribution function in the infrared region.

Our approach offers several conceptual advantages over the BLM
scheme. In particular, it is additive and works in cases where the
BLM scale is low. We have emphasized that the value of the BLM scale
alone cannot always be taken as an indicator of the size of
higher-order corrections or the rate of convergence of a perturbative
series. Except in cases where the distribution function is very
narrow, it is better to deduce this information from the distribution
function, which properly takes into account the contribution from all
mass scales. For instance, the distribution function corresponding to
the perturbative series for a quantity which requires a subtraction
of ultraviolet divergences typically gets contributions of opposite
sign, in which case cancellations may occur at one-loop order
resulting in a low value of the BLM scale.

The implementation of our proposal is based on techniques developed
for the analysis of renormalon chains. The distribution function can
be obtained from the integral relations in (\ref{wSrel}) by
calculating first the Borel transform of a perturbative series with
respect to the coupling constant in the limit of large $\beta_0$. We
have demonstrated this for several one- and two-scale problems in
QCD. We find that in many cases the effect of the resummation is
quite significant. As shown in Tables~\ref{tab:5} and \ref{tab:D1},
the difference between the resummed result and the two-loop
approximation is comparable in magnitude to the difference between
the two-loop and the one-loop results, in accordance with the general
behaviour expected for asymptotic series. However, we stress that
whereas the size of the one- and two-loop coefficients are
renormalization-scheme dependent, the result of the resummation is
scale- and scheme-independent. We have associated a scale $\mu_*$
with the resummed series (using a principle value prescription to
regulate the Landau pole in the running coupling constant) and
compared this scale to the BLM scale. Typically, the two scales can
differ by as much as 50\%, which is significant in cases where the
BLM scale is low.

Another aspect in our analysis was to investigate the relative size
of short- and long-distance contributions to the quantities of
interest. Parametrically, long-distance effects are exponentially
small in the coupling constant; they have the form of power
corrections. In practice, however, they can still be sizeable in some
cases. In heavy quark systems, for instance, non-perturbative effects
are often only suppressed by one power of the heavy quark mass. Even
for the bottom quark they can easily reach a level of 10\%, thus
being as large as one-loop perturbative corrections. We have
introduced a hard factorization scale $\lambda=1$ GeV and compared
the short-distance contribution to a model calculation of
long-distance effects. We find that in some cases the long-distance
effects are as big as the short-distance ones. Then no reliable
prediction can be obtained based on perturbation theory alone; it is
necessary to include non-perturbative effects.

In Ref.~\cite{part2}, we generalize our resummation procedure to the
description of cross sections and inclusive decay rates. In the
calculation of radiative corrections both virtual and real gluons
will have to be considered, and only the sum of their contributions
is infrared finite \cite{Kino,LeeN}. Clearly, in such a situation one
has to generalize the idea of performing a one-loop calculation with
a running coupling constant, which was the motivation for our
resummation. As a consequence, the ``linear'' form of the integral
representation given in (\ref{Sappr}) will be replaced by
``non-linear'' representations, in which instead of the coupling
constant $\alpha_s$ there appears a function of the coupling
constant.

\bigskip
While this paper was in writing, I became aware of a preprint by
Beneke and Braun \cite{BBnew}, who propose the same generalization of
the BLM scheme. Their approach is similar in spirit to the one
presented here, although the formalism differs in technical details.
I am grateful to the authors for making their results available to me
prior to publication.

\subsection*{Acknowledgements}

It is a pleasure to thank G. Altarelli, P. Ball, J. Ellis, B. Gavela,
M. Jamin, A. Kataev, G.~Martinelli, P. Nason, O. P\`ene and A. Pich
for useful discussions.

\end{document}